\documentclass[reprint,superscriptaddress,longbibliography, amsmath,amssymb,aps,prb]{revtex4-1}

\usepackage{amsmath}
\usepackage[colorlinks=true,citecolor=blue]{hyperref}
\usepackage{graphicx}% Include figure files
\usepackage{bm}% bold greek math
\usepackage[dvipsnames]{xcolor}
\usepackage[normalem]{ulem}
\usepackage{gensymb}% Special symbols
\UseRawInputEncoding
\begin{document}

\preprint{APS/123-QED}

\title{Antiferromagnet-mediated interlayer exchange: hybridization versus proximity effect}

\author{D.~M.~Polishchuk}
\affiliation{Nanostructure Physics, Royal Institute of Technology, 10691 Stockholm, Sweden}%
\affiliation{Institute of Magnetism of the NAS of Ukraine and MES of Ukraine, 03142 Kyiv, Ukraine}%

\author{Yu.~O.~Tykhonenko-Polishchuk}
\affiliation{Nanostructure Physics, Royal Institute of Technology, 10691 Stockholm, Sweden}%
\affiliation{Institute of Magnetism of the NAS of Ukraine and MES of Ukraine, 03142 Kyiv, Ukraine}%

\author{Ya.~M.~Lytvynenko}
\affiliation{Institute of Magnetism of the NAS of Ukraine and MES of Ukraine, 03142 Kyiv, Ukraine}%
\affiliation{Institut f\"ur Physik, Johannes Gutenberg Universit\"at Mainz, D-55128 Mainz, Germany}%

\author{A.~M.~Rostas}
\affiliation{National Institute of Materials Physics, 077125 Bucharest-Magurele, Romania}%

\author{V.~Kuncser}
\affiliation{National Institute of Materials Physics, 077125 Bucharest-Magurele, Romania}%

\author{A.~F.~Kravets}
\affiliation{Nanostructure Physics, Royal Institute of Technology, 10691 Stockholm, Sweden}%
\affiliation{Institute of Magnetism of the NAS of Ukraine and MES of Ukraine, 03142 Kyiv, Ukraine}%

\author{A.~I.~Tovstolytkin}
\affiliation{Institute of Magnetism of the NAS of Ukraine and MES of Ukraine, 03142 Kyiv, Ukraine}%
\affiliation{Faculty of Radiophysics, Electronics and Computer Systems, Taras Shevchenko National University of Kyiv, 64/13 Volodymyrska Str., 01601 Kyiv, Ukraine}%

\author{O.~V.~Gomonay}
\affiliation{Institut f\"ur Physik, Johannes Gutenberg Universit\"at Mainz, D-55128 Mainz, Germany}%

\author{V.~Korenivski}%
\affiliation{Nanostructure Physics, Royal Institute of Technology, 10691 Stockholm, Sweden}%
\email{vk@kth.se.}

\begin{abstract}
We investigate the interlayer coupling between two thin ferromagnetic (F) films mediated by an antiferromagnetic (AF) spacer in F*/AF/F trilayers and show how it transitions between different regimes on changing the AF thickness. Employing layer-selective Kerr magnetometry and ferromagnetic-resonance techniques in a complementary manner enables us to distinguish between three functionally distinct regimes of such ferromagnetic interlayer coupling. The F layers are found to be individually and independently exchange-biased for thick FeMn spacers -- the first regime of no interlayer F-F* coupling. F-F* coupling appears on decreasing the FeMn thickness below 9~nm. In this second regime found in structures with 6.0-9.0~nm thick FeMn spacers, the interlayer coupling exists only in a finite temperature interval just below the effective Néel temperature of the spacer, which is due to magnon-mediated exchange through the thermally softened antiferromagnetic spacer, vanishing at lower temperatures. The third regime, with FeMn thinner than 4 nm, is characterized by a much stronger interlayer coupling in the entire temperature interval, which is attributed to a magnetic-proximity induced ferromagnetic exchange. These experimental results, spanning the key geometrical parameters and thermal regimes of the F*/AF/F nanostructure, complemented by a comprehensive theoretical analysis, should broaden the understanding of the interlayer exchange in magnetic multilayers and potentially be useful for applications in spin-thermionics.
\end{abstract}

\maketitle

\section{Introduction} 

Antiferromagnetic (AF) spintronics is an emerging field of research, where focus is on electrical, optical, and other means of controlling the AF order parameter and its utility in electronic devices. AF order is highly stable against perturbations by magnetic fields, produces no stray fields, displays ultrafast (THz) spin dynamics, and can generate large magnetotransport effects~\cite{Jungfleisch2018}. Several new effects have recently been found in AF materials, such as the tunnel anisotropic magnetoresistance~\cite{Wang2014, Marti2014, Kriegner2016}, anisotropic magnetoresistance~\cite{Bodnar2020, Shick2010, Wang2012, Park2011}, spin Seebeck~\cite{Wu2016, Rezende2016}, inverse spin Hall~\cite{Mendes2014, Qu2015}, inverse spin galvanic, and other effects~\cite{Reichlova2015, Zhang2014}. As a result, nanostructures incorporating antiferromagnets have become the topic of intense research for spintronic applications.

Antiferromagnets have traditionally been used in nanodevices for exchange-biasing ferromagnetic elements used in memory, sensing, and other applications. Exchange bias occurs in, e.g., ferromagnet/antiferromagnet (FM/AF) bilayers due to the interlayer coupling via the interfacial spins of the FM and AF coming in direct contact. It is informative to note for the discussion to follow that such type of interlayer coupling is necessarily weaker than the intralayer exchange coupling (within FM and AF) due to typically inhomogeneous spin distributions and some orientational frustration at the FM/AF interfaces originating from the competing FM- vs AF-exchange ordering, superposed surface roughness, interdiffusion, structural defects (e.g., grain boundaries in polycrystalline films), etc.~\cite{Nogues1999}. In addition to the rather passive role in exchange-biasing, the recent years have seen antiferromagnets used as active components in spin transport nanodevices~\cite{Wadley2016, Jungwirth2016, Jungwirth2018,Zhou2020}.

\begin{figure}[b]%[t,b,c,h]
\includegraphics[width=8.5 cm]{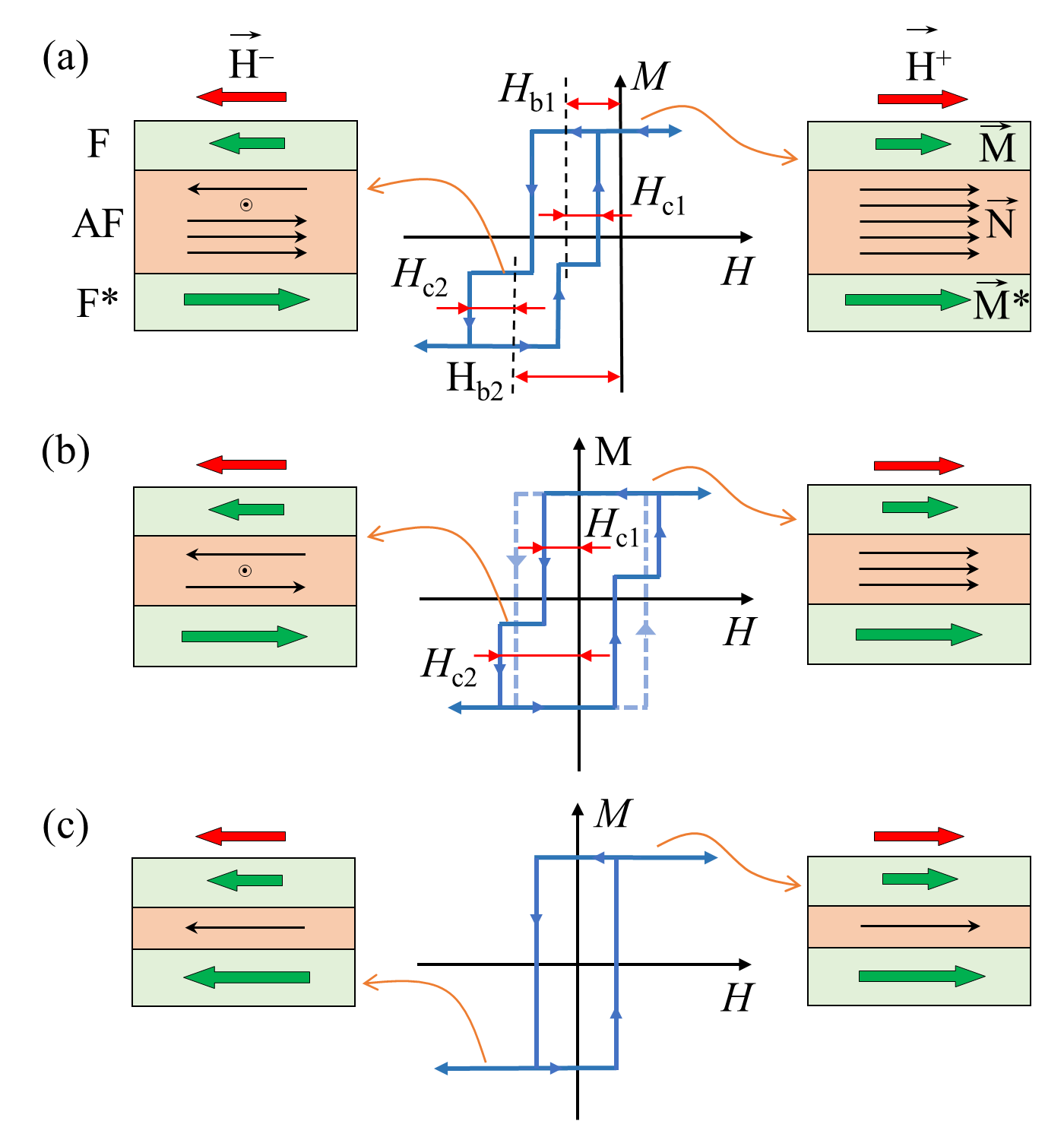}%{manMagnonExchange/fig_frequencies3.jpg}%
\caption{Schematic of exchange-coupled trilayer of type ``hard-ferromagnet/antiferromagnet/ soft-ferromagnet'' (F*/AF/F), with red arrows corresponding to external field, green arrows to F-layers' magnetic moments (of different magnitude; in parallel and antiparallel states), and black arrows to antiferromagnetic vector N within spacer: (a) case of no interlayer coupling -- strong-AF limit, thick AF spacer; (b) finite interlayer coupling at near $T_\mathrm{N}$ (dashed) and no coupling at lower $T$ (solid) -- weak AF in spacer of intermediate thickness; and (c) strong interlayer coupling -- vanishing AF order in ultra-thin spacer.}
\label{fig_1}
\end{figure}

To analyze the possible cases of AF-mediated exchange coupling between FM layers [Fig.~\ref{fig_1}(a-c)], one needs to consider the effect of exchange bias in more detail. Although the full microscopic description of this phenomenon is still under development~\cite{Stamps2000, Kiwi2001}, a good starting point is to picture the AF spins as aligned by exchange preferentially in parallel to the FM spins at the FM-AF interface, which is typically achieved during cooling the structure in a magnetic field~\cite{Nogues1999, Stamps2000, Bobo2004, Berkowitz1999}. This model has been experimentally confirmed for most metallic FM-AF interfaces~\cite{Mohanty2013, Antel1999}, whereas for FM/oxide-AF interfaces the spin orientations in FM and AF are often orthogonal~\cite{OHandley2000}. 

In a FM/AF bilayer, compared to a free-standing film, the FM layer is under an additional torque exerted by the exchange coupling to the AF. One can distinguish two limiting cases, depending on the strength of the intrinsic AF-ordering and/or anisotropy in the AF layer. A \textit{strong} AF (strong AF ordering and/or anisotropy) acts to field-offset the hysteresis loop of the FM layer [Fig.~\ref{fig_1}(a)], whereas a \textit{weak} AF often enhances the F-layer coercivity without offsetting its hysteresis loop [Fig.~\ref{fig_1}(b)]~\cite{Nogues2005}.

For a proper evaluation of the exchange bias in the strong-AF case, the coupling energy for a given interface, $E = H_b M_s V_\mathrm{FM}$, should be used instead of the exchange-bias field, $H_b$~\cite{Nogues2005}. Here $M_s$ is the saturation magnetization and $V_\mathrm{FM}$ -- the volume of the ferromagnet. As an example, in F*/AF/F trilayers where the coupling energy is same for the two interfaces, $H_b$ depends inversely on $M_s$, as illustrated in in Fig.~\ref{fig_1}(a). 

%==================== Figure 1 ====================
%==================================================

In contrast, in the weak-AF case, it is energetically favorable for the AF’s surface spins to follow the rotation of the FM magnetization in a reversing magnetic field (rather than pinning its direction, up to the exchange-offset field). The resulting, often irreversible spin perturbation in the soft AF layer and the associated extra energy is the reason for the enhanced coercivity, manifest as a broader FM hysteresis loop, with a larger magnetic field, $H_c$, required to reverse the FM magnetization, for both positive and negative branches. In this case, the loop can display no field offset~\cite{Nogues2005}; see Fig.~\ref{fig_1}(b).

Interestingly and highly useful for studying the built-in energetics, increasing the system’s temperature toward the AF’s Néel point, $T_\mathrm{N}$, weakens the AF and can invoke a transition between the above two cases of the AF-mediated interlayer coupling, manifest as a concomitant reduction of $H_b$ and enhancement of $H_c$~\cite{Nogues1999, Leighton2002}.

The interfacial region of the perturbed spin distribution resides mainly in the AF layer and vanishes into AF over a characteristic length ($\lambda_\mathrm{in}$) of the order of a few nanometers. A decrease in the thickness of the AF spacer layer to $t \lesssim 2 \lambda_\mathrm{in}$ weakens the overall strength of the AF-ordering and leads to a significant reduction of the effective Néel temperature~\cite{Merodio2014} of the spacer. This effect of the penetrating FM-exchange field, initially observed in FM systems~\cite{Hernando1995, Navarro1996}, favoring parallel alignment of the AF spins and resulting in a non-zero induced AF magnetization, is known as the magnetic \textit{proximity} effect~\cite{Lenz2007}. 

Among metallic antiferromagnets, FeMn (Fe$_{50}$Mn$_{50}$) is a rather unique material. Its lattice parameter is close to that of a number of iron-based FM alloys, which makes it possible to fabricate various nearly stress-free FM/AF multilayers, a typical example of which is Py/FeMn~\cite{Saglam2016, Ekholm2011} (Py = Ni$_{81}$Fe$_{19}$, Permalloy). Its Néel temperature ($T_\mathrm{N}$) is thickness-dependent; it decreases with a progressive decrease in the AF layer thickness to a few nanometers from about 150~$\degree$C to/below room temperature. This flexibility in the operating temperature range makes FeMn-based nanostructures attractive for applications as well as for exploring various basic-physics aspects of AF-based multilayers. 

In this work, we investigate the interlayer coupling in hard-FM/AF/soft-FM sandwiches, acting between the outer FM layers via the AF spacer and find three characteristic regimes in the system’s behavior depending on the AF spacer thickness. The first regime is bulk-like where the FeMn spacer is relatively thick and strongly AF-ordered. Here, the outer ferrromagnetic layers are strongly and independently exchange-pinned at all temperatures below the Néel point of the spacer. The second regime is where the spacer is too thin to display any significant AF-order and is partially FM-polarized by the outer F-layers. In this second case, the outer FM layers are strongly coupled by a direct FM exchange interaction, forming effectively a single FM layer, at least when probed magneto-statically. The third regime is intermediate, in which the spacer displays AF-like or FM-like character depending on the proximity to $T_N$. The interlayer coupling in this third regime is feromagnetic in a narrow temperature range just below the Néel point and has been recently discussed in detail~\cite{Polishchuk2021}. The second, strong-proximity regime is perhaps the least studied to date and receives a special attention in this work, probed by ferromagnetic resonance spectroscopy, which proves to be highly informative specifically in this case.

The temperature and thickness dependence of the interlayer coupling through nominally antiferromagnetic spacers investigated in this work are governed by the competition between the intrinsic AF exchange in the AF spacer and the FM proximity exchange penetrating it from the outer ferromagnetic layers. Our work, carried out on a set of [Fe/Py]/FeMn/Py samples with the FeMn thickness in the range $t = 4-15$~nm, with a particular focus on the ultra-thin spacer limit, is aimed at filling in the gaps in understanding the static and dynamic magnetic properties of AF-based magnetic multilayers. The results obtained should facilitate the development of novel devices for antiferromagnetic spintronics.

%==================== Figure 2 ====================
\begin{figure*}%[t,b,c,h]
\includegraphics[width=17 cm]{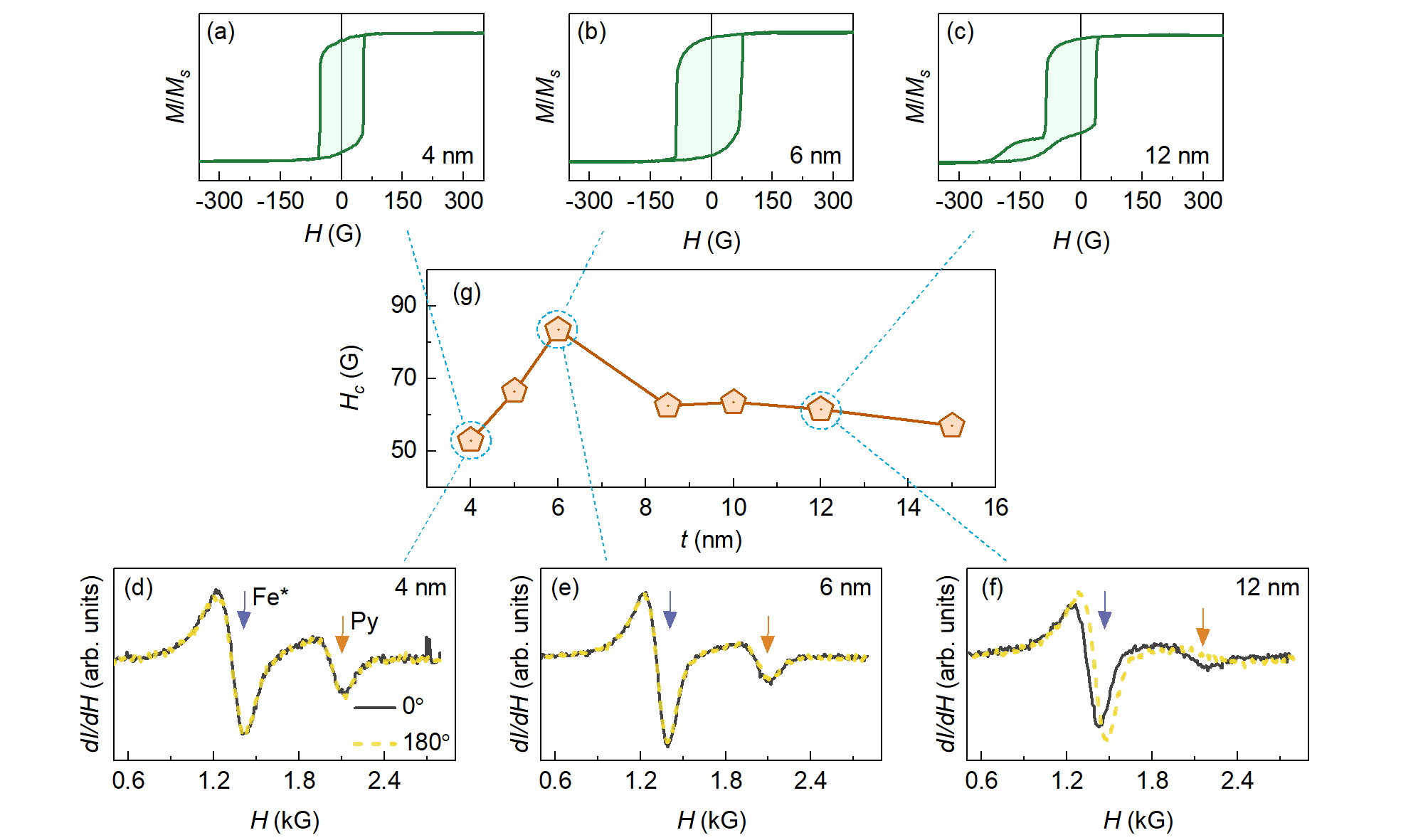}%{manMagnonExchange/fig_frequencies3.jpg}%
\caption{(a)--(c) In-plane hysteresis loops for the Fe*/FeMn($t$)/Py trilayer with $t =$ 4, 6 and 12~nm at room temperature. (g) Thickness dependence of coercivity, $H_c$, and (d)--(f) corresponding broadband FMR spectra at 14~GHz measured at room temperature for two external field orientations with respect to the pinning direction – parallel (0$\degree$) and antiparallel (180$\degree$). Resonance lines from the Fe* ($H_r^\mathrm{Fe*} \approx$ 1.3~kG) and Py ($H_r^\mathrm{Py} \approx$ 2.1~kG) layers are clearly distinguishable.}
\label{fig_2}
\end{figure*}
%==================================================

\section{Samples and methods}

F*/AF/F trilayers used in our experiments had the composition of Cr(5)/[Fe(6)/Py(3)]/FeMn($t$)/Py(5), where the layer thicknesses in nm are given in parenthesis. The Py(5) layer and the [Fe(6)/Py(3)] bilayer are the soft and hard ferromagnetic layers, respectively (hereafter Py and Fe*). The higher-coercive Fe is used to harden the soft Py material in Fe*. The Py(3) sublayer was used to promote the fcc-texture at the [Fe/Py]-FeMn interface for the growth of the FeMn layer with desirable antiferromagnetic properties~\cite{Jungfleisch2018, Wang2014, Antel1999}. Since FeMn displays a strong thickness-dependence due to finite-size effects, we have fabricated a series of trilayers with different thicknesses of the FeMn spacer ($t = 4-15$~nm). We also prepared reference Cr(5)/[Fe(6)/Py(3)]/FeMn($t$)/Al(4) ($t = 4-15$~nm) (hereafter -- Fe*/FeMn($t$)) and Ti(5)/Cu(20)/Py(7)/Ti(5) (hereafter -- Cu/Py) samples. The multilayers were deposited by DC magnetron sputtering (Orion, AJA Intern.) at room temperature. To induce a preferred magnetization direction, the samples were deposited and subsequently annealed at 250~$\degree$C in the presence of a saturating dc magnetic field applied in-plane.

The static magnetic properties were characterized using polarization-modulated magneto-optical Kerr effect (PM-MOKE) measurements in the longitudinal configuration in the temperature range of 100--450~K using a home-built setup equipped with an optical cryostat (Oxford Instruments). The polarization modulation enabled us to obtain magnetic material-selective MOKE signals by measuring the Kerr ellipticity (1$^\mathrm{st}$ harmonic) and Kerr rotation (2$^\mathrm{nd}$ harmonic)~\cite{Nederpel1985,Polisetty2008}. The measurements of microwave-cavity ferromagnetic resonance (FMR) were carried out in the temperature range of 100--300~K at a fixed frequency of 9.44~GHz using X-band ESR spectrometer model EMX-plus (Bruker Inc.). The room-temperature magnetic properties were characterized using a microstrip-based FMR setup and a vibrating sample magnetometer (Lakeshore Cryogenics). More details on the fabrication and characterization of magnetic multilayered structures of the type discussed in this paper can be found in our earlier publications~\cite{Kravets_2014, Kravets_2015, Kravets_2016}. 

\section{Results and discussion}

\subsection{Magnetic Properties}

The evolution in the magnetic properties of Fe*/FeMn($t$)/Py upon varying the thickness of the AF spacer is illustrated by Fig.~\ref{fig_2}(a-g) with the results of VSM and broadband FMR measurements at 300~K.
 
Panels (a)--(c) of Fig.~\ref{fig_2} illustrate the transformation of the shape of the hysteresis loops with changing the FeMn thickness. For relatively thick FeMn ($t >$ 6~nm), the hysteresis loops consist of two contributions with distinctly different values of $H_b$ and $H_c$. As explained above, of the two FM layers adjacent to the AF spacer, the layer with higher magnetization is expected to have a smaller $H_b$. We therefore attribute the hysteresis loop with the smaller $H_b$ to the Fe* layer. The other, strongly field-offset hysteresis loop belongs to the Py layer. This behavior agrees with the illustration in Fig.~\ref{fig_1}(a) of the strong-AF limit, characterized by strong exchange-bias.

For the structure with a 6-nm FeMn spacer, one can observe only one contribution to the hysteresis loop at room temperature [Fig.~\ref{fig_2}(b)], with $H_b =$ 0 and enhanced coercivity, which agrees with the situation illustrated in Fig.~\ref{fig_1}(b) (dashed loop for near $T_\mathrm{N}$) -- weak AF, negligible exchange bias, enhanced coercivity. 

Further decrease in the FeMn thickness does not change the shape of the hysteresis loop but leads to a decrease in coercivity. It appears that the coupling between the Fe* and Py layers becomes stronger (to be verified by FMR), which represents the regime illustrated in Fig. 1(c) -- interlayer coupling via the FeMn spacer ferromagnetically polarized due to the magnetic proximity effect from the adjacent Fe* and Py layers.

The broadband-FMR data [insets (d)--(f) to Fig.~\ref{fig_2}] are consistent with the above picture. For the samples with thicker AF spacers, exemplified for $t =$ 12~nm in Fig.~\ref{fig_2}(f), the presence of exchange bias is clearly evident for both ferromagnetic layers: the resonance field, $H_r$, is significantly different when the external field applied along or against the nominal direction of the exchange pinning. This indicates that the magnetic field applied during the FMR-measurements is not able to change the magnetic configuration of the antiferromagnet (strong AF). Further, it can be seen that the exchange bias for the Py layer is higher than that for the Fe* layer.

For $t \leq $ 6~nm, the resonance fields for the Fe* and Py layers are independent of the field direction, which agrees with the VSM data and indicates that the system does not display exchange bias (no field offset).

\subsection{Layer-Selective MOKE Measurements}

\begin{figure*}%[t,b,c,h]
\includegraphics[width=17 cm]{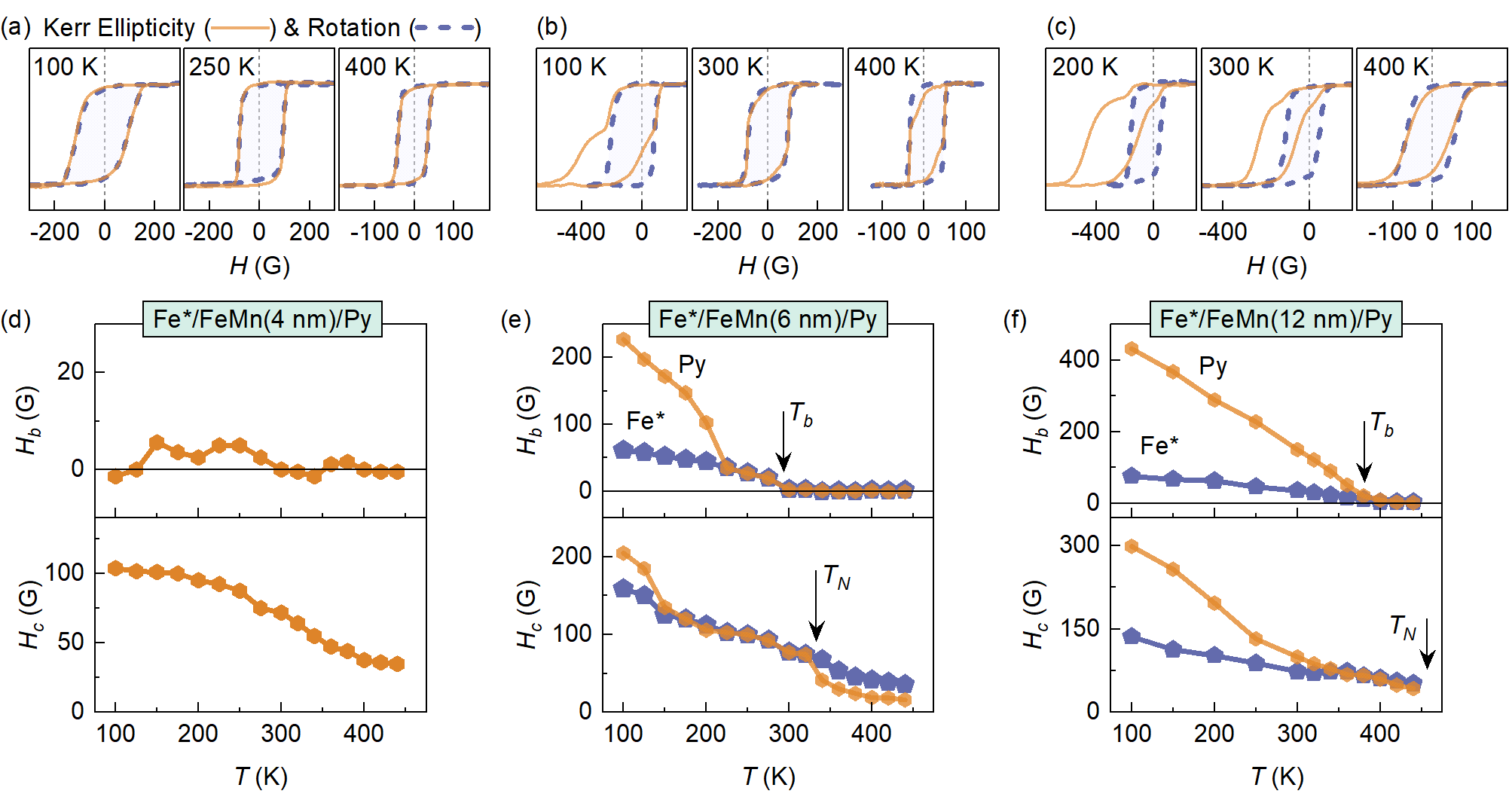}%{manMagnonExchange/fig_frequencies3.jpg}%
\caption{(a--b) First- and second-harmonic MOKE hysteresis loops for Fe*/FeMn($t$)/Py trilayer with $t =$ 4, 6, and 12~nm, in the temperature range 100--400~K. (d--f) Temperature dependence of the exchange bias field, $H_b$, and coercivity field, $H_c$, for the soft Py and hard Fe* layers for the samples with $t =$ 4, 6, and 12~nm, respectively.}
\label{fig_3}
\end{figure*}

The polarization-modulation technique of measuring MOKE allows separating the loops of Fe* and Py layers due to the difference in the magnetooptical properties of Ni and Fe~\cite{Polishchuk2018,Polishchuk2021}. MOKE loops at selected temperatures for the structures with different FeMn thicknesses are shown in panels (a)--(d) of Fig.~\ref{fig_3}. Kerr ellipticity reflects the reversal of the two FM layers, while Kerr rotation is much more sensitive to Fe* vs Py. Overall, the measured MOKE loops are consistent with the results of the VSM measurements.

Analysis of the data shown in Fig.~\ref{fig_3}(a-c) allows one to determine $H_c$ and $H_b$ separately for each FM layer. Calculating the $H_c^\mathrm{Fe*}$ and $H_c^\mathrm{Py}$ coercivity fields using the first field derivative of the MOKE $M$-$H$ loops yields the individual layers’ exchange-bias fields, $H_b^\mathrm{Fe*}$ and $H_b^\mathrm{Py}$. 

The temperature dependence of $H_c$ and $H_b$ for $t =$ 4, 6, and 12~nm are shown in Fig.~\ref{fig_3}(d-f). All samples with $t >$ 4~nm display significant exchange bias below a certain temperature (blocking temperature, $T_b$). $T_b$ increases on increasing the FeMn thickness, from $\sim$300~K for $t =$ 6~nm to $\sim$390~K for $t =$ 12~nm. At temperatures higher than $T_b$, FeMn, although significantly weakened, preserves AF ordering over an extended temperature range, until $T_\mathrm{N}$ is reached, subsequently undergoing a transition into the paramagnetic state.

%==================== Figure 3 ====================
%==================================================

A quantitative analysis of the MOKE results sheds more light on the characteristics of the interlayer coupling in the system with the AF spacer varied in thickness and yields the following three qualitatively different regimes.

(i) $t >$ 6~nm; Fig.~\ref{fig_3}(f). At high temperatures ($T > T_\mathrm{N}$), where the spacer is paramagnetic, the two FM layers are exchange decoupled. The MOKE data show two separate loops, which have no offset along the field axis ($H_b = 0$), have different coercivity, and correspond to essentially free soft-Py and hard-Fe* layers ($H_c^\mathrm{Py} < H_c^\mathrm{Fe*}$). 

Lowering temperature below $T_b$ produces a non-zero exchange bias in both Fe* and Py, which increases with further lowering $T$. $H_b$ for Fe* is smaller than that for Py, which is expected since, for the same coupling energy at the AF-FM interface, $H_b$ is inversely proportional to the FM-layer magnetic moment.

The behavior of coercivity is more nuanced. The coercivity of both FM layers increases with decreasing temperature, but after crossing $T_b$, the change in $H_c^\mathrm{Py}$ becomes much more pronounced than that in $H_c^\mathrm{Fe*}$. As a result, in the low-temperature region ($T \ll T_b$), $H_c^\mathrm{Py}$ strongly exceeds $H_c^\mathrm{Fe*}$. This may be due to the fact that nominally ultra-soft Py is more susceptible to morphological and spin disorder at the AF interface, compared to the relatively magnetically hard Fe*.

 (ii) $t =$ 6~nm; Fig.~\ref{fig_3}(e). This case is discussed in detail in Ref.~\onlinecite{Polishchuk2021} and briefly treated here for completeness and functional comparison with the focus, high-proximity case (ultra-thin AF). At high ($T > T_\mathrm{N}$) and low ($T \ll T_b$) temperatures, the MOKE hysteresis loops consist of two discernable contributions from the Fe* and Py layers. Within these temperature regions, the behavior of $H_b$ and $H_c$ for both FM layers is similar to the previous case ($t >$ 6~nm). The only difference is that the exchange bias field for both FM layers is smaller, compared to the samples with $t >$ 6~nm. This, along with the reduced $T_b$, agrees with the earlier established characteristic of the system where reduction in the FeMn spacer thickness results in a weaker AF-ordering and lower Néel temperature of the spacer~\cite{Navarro1996, Saglam2016}.
 
In contrast to the above magnetization behavior composed of two superposed hysteresis loops, found at high and low temperatures, a single-loop hysteresis is observed in the intermediate temperature range, just below $T_\mathrm{N}$ ($\sim$225 to $\sim$300~K). In this range, the exchange-bias field is relatively small ($H_b <$ 30~G) but the coercivity is enhanced and exceeds that of all the other samples [see Fig.~\ref{fig_2}(g)]. This unique behavior can be explained~\cite{Polishchuk2021} as due to AF-FM hybridization in the structure and magnon exchange via the weakly antiferromagnetic spacer, which couples the outer layers ferromagnetically. The weak AF-ordering in the spacer in the vicinity of $T_N$ is not sufficient to exchange pin the ferromagnetic layers in a significant way [Fig.~\ref{fig_1}(b)].

As temperature decreases, the single hysteresis loop develops a field offset, which indicates enhanced AF correlations in FeMn and stiffening of the exchange pinning. At a certain temperature, the single loop splits into two minor loops and the behavior transitions in to the first regime above, illustrated in Fig.~\ref{fig_1}(a) -- toward individual pinning of the Fe* and Py layers, with insignificant interlayer FM coupling. 

(iii) $t =$ 4~nm; Fig.~\ref{fig_3}(d). A single loop with a negligible exchange bias ($H_b \approx$ 0) is observed at all temperatures, which is due to vanishing AF order in the thinnest FeMn spacer, dominated instead by the magnetic proximity effect from the Fe* and Py layers. The outer FM layers are strongly exchange-coupled in this regime, even at high temperatures [Fig.~\ref{fig_3}(a)].

%==================================================

\subsection{Ferromagnetic-Resonance Properties}

%==================== Figure 4 ====================
\begin{figure*}%[t,b,c,h]
\includegraphics[width=17 cm]{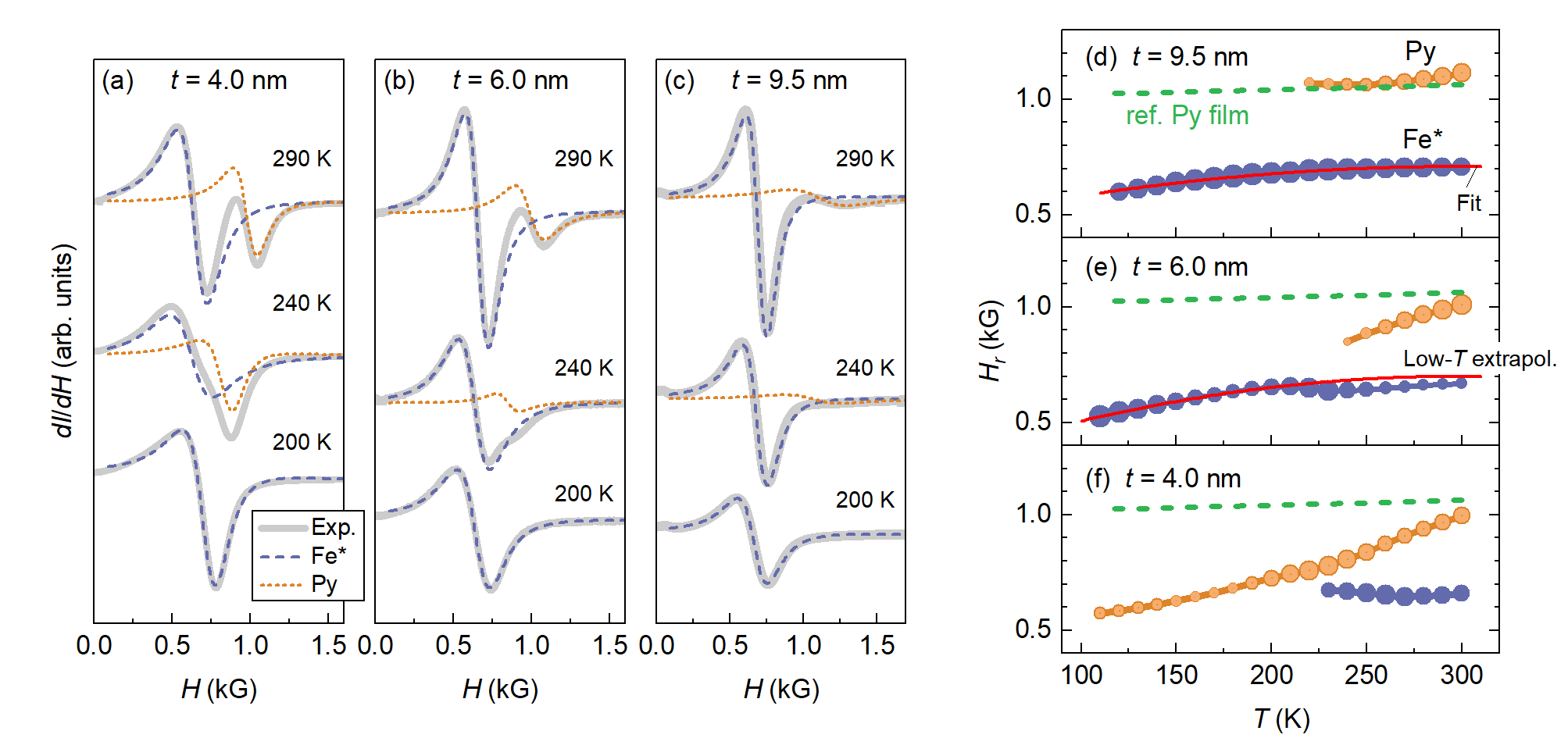}
\caption{Temperature dependent FMR in Fe*/FeMn($t$)/Py trilayers with $t =$ 4, 6, and 9.5~nm. (a--c) In-plane FMR spectra measured perpendicular to the nominal exchange pinning direction; thick solid lines represent experimental data, dashed lines -- components of the fit using the Lorentzian. (d--f) The corresponding Fe* and Py resonance fields as a function of temperature; the size of symbols shows relative changes in line intensity. The dashed line in (d--f) shows the resonance field for the reference Py film. Solid red lines are the fit (d) and low-temperature extrapolation (e) with the same quadratic functional form.}
\label{fig_4}
\end{figure*}
%==================================================

All Fe*/FeMn($t$)/Py structures reveal two well-defined resonance lines (from Fe* and Py) at room temperature, as seen in the broad-band FMR spectra shown in Fig.~\ref{fig_2}(d-f). The second technique we use -- cavity-FMR at a fixed frequency of $\sim$10~GHz -- reproduces these lines with much better signal-to-noise ratio and allows us to investigate the temperature evolution of the FMR spectra and compare the temperature dependence of the resonance fields for Fe* and Py in the samples with different $t$; Fig.~\ref{fig_4}. The FMR spectra were obtained in a configuration where the external field was applied in the film plane and perpendicular to the nominal direction of the exchange pinning. The FMR spectra for a Cu/Py reference sample were used to determine the base-line FMR properties for a free Py layer; marked `ref. Py film' in Fig.~\ref{fig_4}(d-f). 

For the structure with $t =$ 9.5~nm, which represents the case of strong exchange-pinning of Fe* and Py layers, the resonance fields of the two FM layers ($H_r^\mathrm{Fe*}$ and $H_r^\mathrm{Py}$, respectively) demonstrate a weakly temperature-dependent behavior; Fig.~\ref{fig_4}(c,d). The temperature dependence of $H_r^\mathrm{Fe*}$ and $H_r^\mathrm{Py}$ can be explained by the changes in the respective layer’s magnetization and the magnitude of the corresponding exchange-induced anisotropy. It is noteworthy that the temperature dependence of the Py-layer resonance field is essentially that of a single Py film (reference sample; $H_r^\mathrm{refPy}$). The intensity of the Py resonance line is lower for higher exchange coupling (at lower temperature), which is often observed in similar exchange-coupled systems~\cite{Kaya2013, Schmitz2010}.

The temperature-dependent FMR properties of the 6-nm FeMn trilayer are distinctly different from that of the 9.5-nm FeMn structure; cf. Fig.~\ref{fig_4}(d,e). This is particularly visible for the resonance field of the Py line: $H_r^\mathrm{Py}$ deviates from $H_r^\mathrm{refPy}$ and the deviation increases with decreasing temperature -- at 250~K, the difference reaches 16$\%$ of $H_r^\mathrm{refPy}$ ($\approx$170~G). We explain this deviation as due to the substantial interlayer coupling between Fe* and Py, which is also observed in the MOKE study as a single $M-H$ loop in the temperature interval of $225-320$~K. What is impossible using the MOKE data, from this difference in the $H_r$ we can estimate the strength of the interlayer coupling, $J^*_\mathrm{IC}$, which can reach $\sim$0.05~erg/cm$^2$. It should be noted, however, that the actual $J_\mathrm{IC}$ should be somewhat lower than the estimated $J^*_\mathrm{IC}$ because the latter does not exclude an additional factor -- so-called rotatable magnetic anisotropy that is often present in such ultra-thin FM-AF multilayers~\cite{Polishchuk2021a}.

The intensity of the Py line becomes too weak below $T \approx$ 240~K for reliable determination of its position. Along with the non-zero $H_b$ from the MOKE study, the vanishing Py-line intensity additionally confirms the substantial strengthening of the exchange pinning with decreasing temperature below $\sim$250~K. This is the result of a magnetic phase transition in the FeMn spacer, when its AF order becomes strong enough to exchange-pin the outer Fe* and Py layers individually. Consequently, there is no interlayer coupling between Fe* and Py at low temperatures ($T <$ 225~K).

The Fe* line, on the contrary, remains highly intensive in the whole temperature interval [Fig.~\ref{fig_4}(b)], which can be explained by the $\sim$3 times larger magnetic moment of the Fe* layer than the one of Py. For the same reason, the position of the Fe* line is less sensitive to the interlayer coupling; cf. Fig.~\ref{fig_4}(e). Despite that, $H_r^\mathrm{Fe*}$-vs-$T$ shows well defined deviation above 225~K from the extrapolation of the low-temperature trend (solid red line in Fig.~\ref{fig_4}(e)). Importantly, the low-temperature part of $H_r^\mathrm{Fe*}$ ($T <$ 225~K) has the same temperature-dependent trend as the one for the 9.5-nm structure. This implies that the low-temperature properties of the 6-nm structure are governed by the strong exchange pinning akin to that of the 9.5-nm structure.

The FMR study of the 4-nm FeMn structure proves to be much more informative than the MOKE data regarding the strength and nature of the interlayer coupling in the structure. When the MOKE yields a single hysteresis loop with no exchange offset and no changes in its shape in the whole temperature interval of $100-460$~K [Fig.~\ref{fig_3}(a)], the FMR reveals substantial temperature-induced variations in the dynamic properties of the structure; see Fig.~\ref{fig_4}(a,f). Similar to the 6-nm FeMn structure, there are two well defined resonance lines at higher temperatures ($T \geq$ 230~K), where $H_r^\mathrm{Py}$ considerably deviates from $H_r^\mathrm{refPy}$ (by 21$\%$ at 250~K). What is different for the 4-nm FeMn sample is that the two lines clearly merge into a single mode below 230~K. The presence of this low-temperature single mode and the absence of exchange pinning indicate a regime of the interlayer coupling through the 4-nm FeMn that is qualitatively different from those revealed for the structures with the thicker FeMn spacers. We explain this regime as caused by the strong magnetic proximity effect, as discussed below.

The difference in the resonance fields $H_r^\mathrm{Py}$ and $H_r^\mathrm{refPy}$ can give the strength of the FeMn-mediated interlayer coupling for the structures with $t =$ 4~nm and 6~nm. According to the standard FMR formalism~\cite{Kittel1948}, the interlayer coupling affects the FMR behavior of the Py layer by adding the effective field of interlayer coupling, $H_\mathrm{IC}$, to the total effective field $H_\mathrm{eff}$ acting on the spins in Py. Assuming the absence of in-plane magnetic anisotropy, $H_\mathrm{eff} = (H + H_\mathrm{IC})$, where $H$ is the external magnetic field. Then, the FMR frequency for the in-plane orientation of $H$ reads $f_\mathrm{res} = (\gamma/2\pi)[H_\mathrm{eff}(H_\mathrm{eff} + 4\pi M_s)]^{1/2}$, where $\gamma$ is the gyromagnetic ratio, $4\pi M_s$ -- the demagnetizing field of a thin film. The difference between $H_r^\mathrm{refPy}$ of the free Py ($H_\mathrm{eff} = H$) and $H_r^\mathrm{Py}$ of the exchange coupled Py ($H_\mathrm{eff} = H + H_\mathrm{IC}$), gives $H_\mathrm{IC}$ that can be converted into the constant of interlayer coupling as $J_\mathrm{IC} = H_\mathrm{IC}M_s t_\mathrm{Py}$, where $t_\mathrm{Py}$ is the thickness of Py. Numerically, $J_\mathrm{IC}$ for the 4~nm and 6~nm samples are comparable in the vicinity of the Néel transition (of the order of 0.1~erg/cm$^2$) and diverge at low temperature. $J_\mathrm{IC}$ vanishes for the 6~nm spacer and increases several fold for the 4~nm spacer, directly reflecting the behavior of $H_\mathrm{IC}$ extracted from the FMR data.  

\begin{figure}%[t,b,c,h]
\includegraphics[width=8.5 cm]{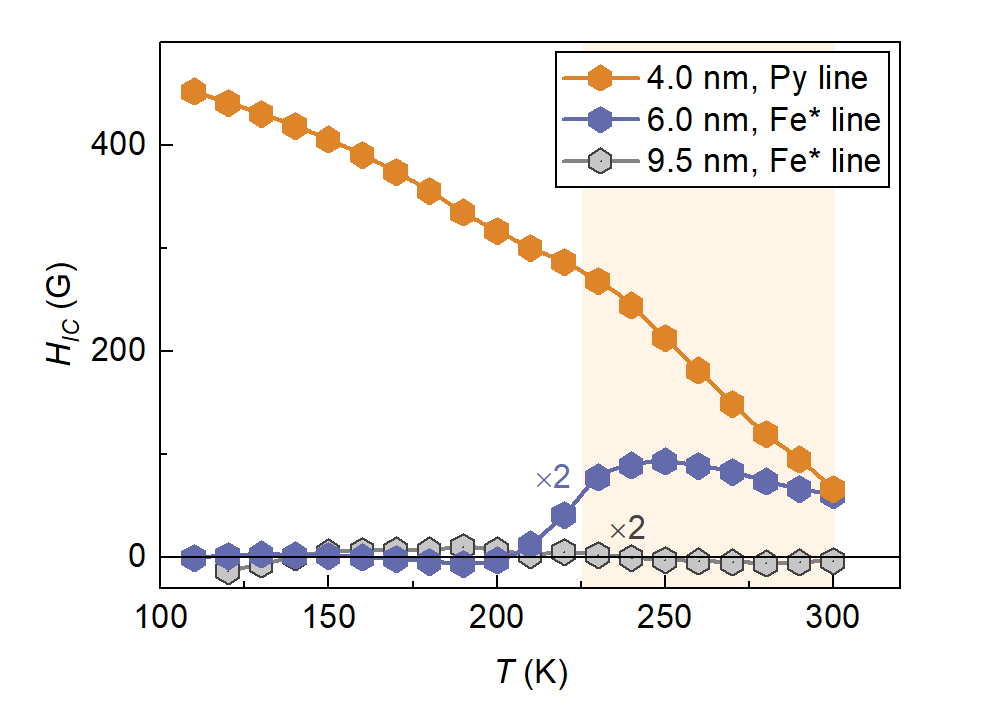}
\caption{Effective field of interlayer coupling, $H_\mathrm{IC}$, determined as difference in resonance fields for Py layer, $H_r^\mathrm{Py}$, and reference Py film, $H_r^\mathrm{refPy}$, for $t =$ 4~nm, and as the deviation from the low-temperature extrapolation of $H_r^\mathrm{Fe*}(T)$ for $t =$ 6,9~nm.}
\label{fig_5}
\end{figure}

Since the Py line vanishes in intensity below 240~K for the sample with $t =$ 6~nm, the Fe* line can be used instead at low temperatures. As seen in Fig.~\ref{fig_4}(e), the extrapolation of the $H_r^\mathrm{Fe*}(T)$ yields a clear deviation at $T >$ 220~K. This extrapolation uses the same quadratic functional form that perfectly fits the $H_r^\mathrm{Fe*}(T)$ of the 9.5-nm FeMn sample in the whole temperature interval; Fig.~\ref{fig_4}(d). The deviation from the extrapolation for the 6-nm sample is proportional to $H_\mathrm{IC}$ and indicates the presence of the interlayer coupling between Fe* and Py in the finite temperature interval, $T \geq$ 220~K, shown in Fig.~\ref{fig_5}. Owing to the 3-times larger magnetic moment of Fe*, the respective $H_\mathrm{IC}$ is numerically smaller than that determined for the Py line.

The interlayer coupling for the 4-nm FeMn structure at low temperatures can still be traced using the Py line, which in fact merges with the Fe* line at $T <$ 230~K; cf. Fig.~\ref{fig_4}(f). The merging indicates much stronger coupling present down to 100~K, which is in stark contrast to the behavior found for the 6-nm sample.

%==================== Figure 5 ====================
%==================================================

\subsection{Three regimes of interlayer coupling in F/AF/F}

The observed interlayer coupling through the nominally antiferromagnetic FeMn spacer and its thickness- and temperature dependence can be explained by an interplay between finite-size and magnetic proximity effects on the one hand and the intrinsic AF order in the spacer on the other. It is known that the Néel temperature of ultra-thin FeMn films is suppressed to below room temperature~\cite{Merodio2014} owing to the finite-size effect -- the effective intrinsic AF exchange is weakened by the absence of the nearest-neighbor AF order for the surface spins. The AF properties of the thinnest FeMn spacers in our structures are further suppressed by the proximity of the outer Fe* and Py layers, which magnetize the interface regions of the spacer and thereby counteract AF ordering, facilitating \textit{ferromagnetic} interlayer coupling.

Our Fe*/FeMn(4~nm)/Py trilayers show no Néel transition in the entire temperature interval, as indicated by a single hysteresis loop measured in the entire experimental range, the non-zero induced magnetization in the AF spacer (not shown), as well as the continuously increasing interlayer coupling strength toward the lowest temperatures. All indications are that AF spacers of sub-critical thickness (below 6~nm of FeMn) behave as weak ferromagnets as regards the interlayer exchange. The characteristic penetration depth of the proximity effect for ferromagnetically enclosed FeMn can thus be estimated as about 2 nm for our structures. It should be noted, however, that the effect is non-linear versus the AF thickness, so the penetration depth is likely a function of the multilayer geometry as well as its material parameters (details go beyond the scope of this paper). 

The limit of thick AF spacers, 9~nm and above, is straightforward to interpret. The finite-size effect is relatively weak, such that the AF-ordering and the Néel temperature are essentially those found in the bulk FeMn alloy. The strong AF-ordering counteracts the proximity effect much more efficiently with the result of fully suppressing the interlayer exchange coupling. This is manifest via, in particular, the measured bulk-like blocking temperature as well as the mutually independent exchange-bias fields of the two outer ferromagnetic layers. 

The case of the critical spacer thickness of about 6~nm is rather unique and was analyzed in great detail in~\cite{Polishchuk2021}. Here, the F-proximity helped by the finite-size effect is able to counteract and penetrate the spacer only in the vicinity of the Nèel temperature, where the AF-order is weak. Just below $T_N$, the magnon modes in the AF spacer hybridize with those in the ferromagnetic layers, which results in a magnon mediated interlayer coupling. As the temperature is lowered and the AF order is strengthened, the proximity-induced magnetization is 'squeezed out', and the interlayer coupling is again fully suppressed. 

The qualitative difference in the functional form of the AF-mediated interlayer coupling between the two most interesting cases of critical and sub-critical spacer thickness is most clearly visible in Fig.~\ref{fig_5}, where the non-monotonic, thermally-gated dependence for the 6-nm spacer transforms into a continuously rising trend for the 4-nm spacer. These results should be useful in tuning interlayer coupling in AF-based multilayers.  

\subsection{Theoretical considerations}

\begin{figure}%[htbp]
\begin{center}
\includegraphics[width=8.5 cm]{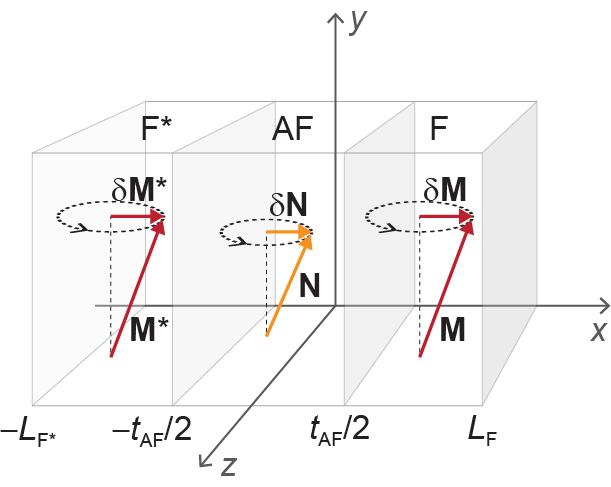}
\caption{Schematic of F*/AF/F trilayer.}
\label{fig_Temp}
\end{center}
\end{figure}

To interpret the observed behavior, functionally different in different temperature intervals, we develop an analytical spin-dynamic model of the trilayer. Its structure consists of two ferromagnetic layers, F* and F, separated by an AF layer of thickness $t_\mathrm{AF}$; see Fig.~\ref{fig_Temp}. The thicknesses of the ferromagnetic layers $L_\mathrm{F}, L_\mathrm{F*}\gg t_\mathrm{AF}$, but much smaller than the domain wall width in F*(F), so the spin distribution is homogeneous within the ferromagnets. The antiferromagnetic layer is a collinear antiferromagnet, whose  magnetic state is described by the N\'eel vector, $\mathbf{N}$, and magnetization, $\mathbf{M}_\mathrm{AF}$. The ferromagnetic layers are characterized by magnetization vectors $\mathbf{M}^*$ and $\mathbf{M}$. We further take the easy magnetization axes in all three layers to be parallel to $\hat{z}$.

\subsubsection{Equilibrium states of F*/AF/F trilayer}
We start by calculating the possible equilibrium states of the trilayer in the presence  of an external magnetic field, $\mathbf{H}$. For this we minimize the magnetic energy of the system, $W=\int w dx$, where the energy density $w=w_\mathrm{F}+w_\mathrm{F^*}+w_\mathrm{AF}+w_\mathrm{int}$ has four contributions. The ferromagnetic contribution (similar for F and F* layers), 
\begin{equation}\label{eq_ferromagnetic-energy}
    w_\mathrm{F}=-\frac{1}{2} K_\mathrm{an}M_y^2+\frac{1}{2}A\left(\partial_x \mathbf{M}\right)^2+2\pi M_x^2-HM_y,
\end{equation} is nonzero within the ferromagnetic layers, $t_\mathrm{AF}/2<x<L_F$ ( $-L^*_F\le x\le -t_\mathrm{AF}/2$), with $H$ being the value of the external magnetic field parallel to the easy axis. Here $K_\mathrm{an}$ ($K^*_\mathrm{an}$) is magnetic anisotropy and  $A$ ($A^*$) is the exchange stiffness of the F (F*) layer. 

Within the antiferromagnetic layer, the energy density is 
\begin{equation}\label{eq_antiferromagnetic-energy}
w_\mathrm{AF}=\frac{H_\mathrm{ex}}{2M_s}\mathbf{M}^2_\mathrm{AF}-\frac{1}{2} K_\mathrm{AF}N_y^2+\frac{1}{2}A_\mathrm{AF}\left(\partial_x \mathbf{N}\right)^2,
\end{equation}
where $H_\mathrm{ex}$ is the exchange field that keeps antiferromagnetic sublattices antiparallel, $K_\mathrm{AF}$ is the magnetic anisotropy constant, and $A_\mathrm{AF}$ is the exchange stiffness of an antiferromagnet. The Zeeman contribution due to the external magnetic field is implicitly included into the anisotropy value, $K_\mathrm{AF}\rightarrow K_\mathrm{AF}-H^2/(H_\mathrm{ex}M_s)$, however, it can be neglected assuming that $H$ is smaller than the spin-flop field (our experimental case). In most cases (unless specified explicitly) we assume that magnetization of AF is small, so that $|\mathbf{M}_\mathrm{AF}|\ll|\mathbf{N}|$ and  $|\mathbf{N}|=M_s$.

We further assume that the exchange coupling at the F*/AF and AF/F interfaces are parametrized by positive constants $J^*_{b}$ and $J_{b}$ so the interaction energy is
\begin{eqnarray} \label{eq_interaction_dimension}
    w_\mathrm{int}&=&-\xi J^*_{b}(\mathbf{M^*}\cdot\mathbf{N})\delta\left(x+\frac{t_\mathrm{AF}}{2}\right) \nonumber\\ &-&\xi J_{b}(\mathbf{M}\cdot\mathbf{N})\delta\left(x-\frac{t_\mathrm{AF}}{2}\right),
\end{eqnarray}
where $\xi$ is the length-scale characteristic for the interface coupling and $\delta(x)$ denotes the Dirac delta-function. Such interaction favours parallel or antiparallel orientation of all three magnetic vectors at the interfaces. Constants $J^*_{b}$ and $J_{b}$ are related to the effective bias fields, $H^*_b=\xi J_b^*M_s/t_\mathrm{AF}$ and $H_b=\xi J_bM_s/t_\mathrm{AF}$.

The energy of the trilayer is minimal (global minimum in $W$) if all three magnetic vectors $\mathbf{M}^*$, $\mathbf{M}$, and $\mathbf{N}$ are parallel to each other, $\mathbf{M}^*\uparrow\uparrow \mathbf{M}\uparrow\uparrow \mathbf{N}$, and are parallel or antiparallel to the external magnetic field. These states are the only equilibrium states of the system, which are observed experimentally in the weak-AF regime and in the limit of the ultra-thin AF layer (see Fig.~\ref{fig_1}b,c). However, in the strong-AF regime the system's energy ($W$) has a local minimum in a state with the antiparallel alignment of the ferromagnetic moments of the F and F* layers, $\mathbf{M}^*\uparrow\downarrow \mathbf{M}$, with an inhomogeneous spin distribution withing the AF spacer.  We next consider the competition between the parallel and antiparallel states depending on the thickness of the AF layer.

%The energy of this state is higher compared to the collienar state 
\paragraph{Thick AF spacer:} $t_\mathrm{AF}\gg x_\mathrm{DW}$, where  $x_\mathrm{DW}\equiv\sqrt{A_\mathrm{AF}/K_\mathrm{AF}}$ is the thickness of the AF domain wall. In this case, the interaction energy at the interfaces, $w_\mathrm{int}$, is minimized by a parallel alignment of the N\'eel vector and the corresponding ferromagnetic magnetization at each of the interfaces, $\mathbf{N}(-t_\mathrm{AF}/2)\uparrow\uparrow\mathbf{M}^*\uparrow\downarrow \mathbf{M}\uparrow\uparrow\mathbf{N}(t_\mathrm{AF}/2)$. To preserve the continuity of the AF texture, the N\'eel vector rotates through 180$^\circ$ between the two interfaces, forming a domain wall, which is localised close to one of the interfaces. Hence, the reorientation of one ferromagnetic layer is penalised by the energy of the AF domain wall $E_\mathrm{DW}=\sqrt{A_\mathrm{AF}K_\mathrm{AF}}M_s$, which can be considered as a height of the potential barrier between the two states. In the thick-AF limit, $t_\mathrm{AF}\gg x_\mathrm{DW}$, this energy barrier is independent of $t_\mathrm{AF}$. Moreover, the formation of the domain wall at one interface produces no effect on the other, and the two ferromagnetic layers behave independently.

\paragraph{Intermediate AF spacer thickness:} $t_\mathrm{AF}\geq 2x_\mathrm{DW}$. In this case, configuration $\mathbf{N}(-t_\mathrm{AF}/2)\uparrow\uparrow\mathbf{M}^*\uparrow\downarrow \mathbf{M}\uparrow\uparrow\mathbf{N}(t_\mathrm{AF}/2)$ with a 180$^\circ$ domain wall inside the spacer still minimises the interaction energy at the interfaces and the energy barrier is $E_\mathrm{DW}$. However, the domain wall spans the entire space between the ferromagnetic layers and couples $\mathbf{M}^*$ and $\mathbf{M}$, such that they no longer behave independently, as will be discussed later. 

 \paragraph{Thin AF spacer:} $t_\mathrm{AF}\leq 2x_\mathrm{DW}$. An antiparallel alignment of the ferromagnetic moments, $\mathbf{M}^*\uparrow\downarrow \mathbf{M}$, results in a squeezed AF domain wall whose energy $E=E_\mathrm{DW}x\mathrm{DW}/t_\mathrm{AF}$ increases with the decrease of the AF thickness. In other words, the exchange stiffness in AF is large enough to prevent a formation of a domain wall within the AF layer and the only equilibrium states are $\mathbf{M}^*\uparrow\uparrow \mathbf{M}\uparrow\uparrow \mathbf{N}$ (all-parallel up/down). We note that a domain wall in the ferromagnetic layers is disallowed energetically since their thickness is much smaller than the domain wall width in F*/F (experimental case).
 
 \paragraph{Ultrathin AF spacer:} $t_\mathrm{AF}\ll x_\mathrm{DW}$. A further decrease in the AF thickness results in a weakening AF order in the spacer, which lies outside the micromagnetic model discussed above. However, a qualitative analysis is still possible. First, we note that the N\'eel temperature is diminished significantly in ultrathin AF layers~\cite{Lenz2007} and the AF ordering can be fully suppressed ($T_\mathrm{N}=0$) by fluctuations below about seven monolayers in AF thickness ($\sim1.5$ nm for FeMn) \cite{Kuch:PhysRevLett.92.017201}. Even if the exchange in the AF layer remains sufficient to allow long-range ordering within the AF sublattices, it still needs to compete with the exchange coupling at the FM/AF interfaces. For $t_\mathrm{AF}\leq J_bM\xi/H_\mathrm{ex}$, the interfacial exchange coupling produces an effective magnetic field of the order of $\xi (J^*_{b}\mathbf{M^*}+J_{b}\mathbf{M})/t_\mathrm{AF}$~\footnote{In an ultrathin AF spacer, the magnetic sublattices are no longer equivalent due to the presence of the ferromagnetic exchange at the interfaces, so Eq.~\eqref{eq_interaction_dimension} should be modified to include the coupling with both the N\'eel vector and the proximity-induced AF magnetization $\mathbf{M}_\mathrm{AF}$.}, which induces a noticeable canting of the magnetic sublattices in AF and hence nonzero magnetization: 
 \begin{equation}\label{eq_magnetization_ultrathin}
     \mathbf{M}_\mathrm{AF}\propto \frac{\xi}{t_\mathrm{AF}H_\mathrm{ex}}\left(J^*_{b}\mathbf{M^*}+J_{b}\mathbf{M}\right).
 \end{equation}
Such induced magnetization of the AF layer provides an additional coupling between the ferromagnetic layers. As a result, the ferromagnetic layers can couple directly via the proximity-induced magnetization in AF and behave as one under the action of the external field. To conclude, a F*/AF/F trilayer in the limit of ultrathin $t_\mathrm{AF}$ behaves as a ferromagnet whose state is described by a single magnetization vector. The ultrathin limit in our structures is found at $t_\mathrm{AF}\le 4$~nm; our recent study~\cite{Polishchuk:10.1063_5.0133125} on similar Py*/FeMn/Cr/Py multilayers showed an increase of the saturation magnetization below 4~nm in the FeMn spacer thickness attributed to the induced AF magnetization expected in this limit.

Analysis of the equilibrium states points to three possible regimes of the field-induced switching depending on the thickness of AF layer: i) uncoupled, independent switching of the F* and F layers for thick AF spacers; ii) strong coupling and cooperative switching for thin and ultrathin AF spacers; and iii) the intermediate case, in which the transition between the coupled and uncoupled behaviour can be tuned via the magneto-structural parameters of the trilayer. In the next sections we focus on the last \textit{intermediate} case as the most interesting for applications. In particular, we consider the effects of temperature on the degree of ordering in the AF layer and thereby on the spin-dynamic behavior of the trilayer as a whole.

\subsubsection{Energy barrier between the parallel and antiparallel configurations}
\begin{figure}%[htbp]
\begin{center}
\includegraphics[width=8.5 cm]{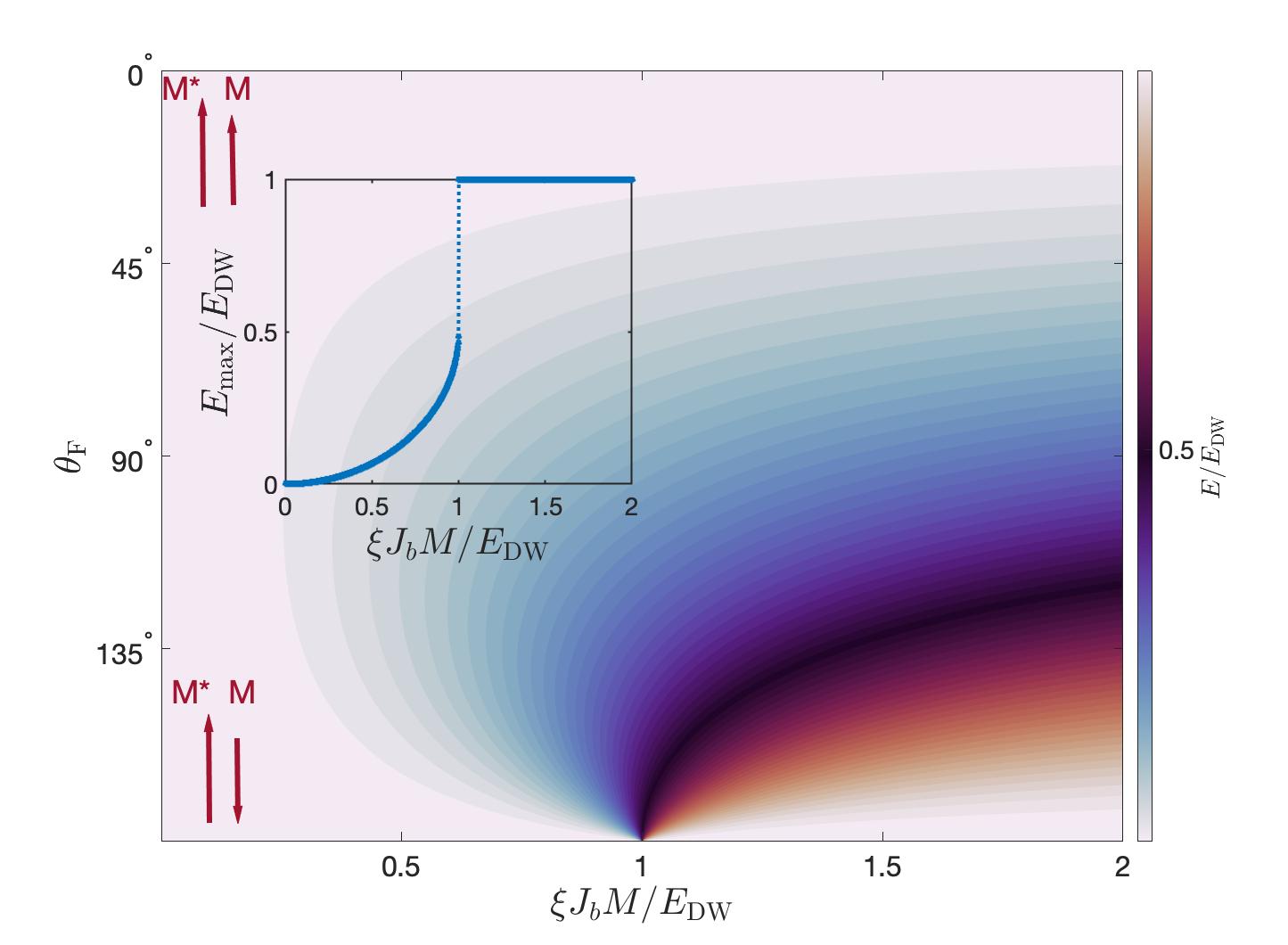}
\caption{Formation of antifparallel F*/AF/F state. Energy of antiferromagnetic texture (color code) depending on angle $\theta_\mathrm{F}$ between $\mathbf{M}$ and $\mathbf{M}^*$ (vertical axis) and ratio between interface coupling and $E_\mathrm{DW}$ (horizontal axis). Inset shows energy maximum as a function of $\xi J_bM/E_\mathrm{DW}$.}
\label{fig_energy_barrier_DW}
\end{center}
\end{figure}

To analyse the transition between the coupled and uncoupled regimes we have calculated the energy barrier between the parallel, $\mathbf{M}^*\uparrow\uparrow \mathbf{M}$, and antiparallel, $\mathbf{M}^*\uparrow\downarrow \mathbf{M}$, configurations. For this we fixed the orientation of $\mathbf{M}^*$ along the easy magnetic axis in one ferromagnetic layer and varied the orientation of $\mathbf{M}$ (parametrized by an angle $\theta_\mathrm{F}$). Figure~\ref{fig_energy_barrier_DW} shows the contribution to the energy of the antiferromagnetic layer associated with the noncollinear alignment of ferromagnetic magnetizations $\mathbf{M}^*$ and $\mathbf{M}$. The energy barrier between the parallel and antiparallel states of the ferromagnets is identified as the energy maximum, $E_\mathrm{max}$, while $\theta_\mathrm{F}$ varies from $0^\circ$ to $180^\circ$.  If coupling at the AF/FM interface is weaker compared to the exchange coupling inside the AF layer, $\xi J_bM\leq E_\mathrm{DW}$, the N\'eel vector is only sightly tilted from the easy direction due to the interface exchange coupling to the ferromagnets,  and the energy barrier is small. In the opposite case, $\xi J_bM\geq E_\mathrm{DW}$, the N\'eel vector rotates through 180$^\circ$ and the energy barrier is equal to the energy of the antiferromagnetic domain wall. Hence, softening of the antiferromagnetic layer impedes formation of the antiparallel $\mathbf{M}^*\uparrow\downarrow \mathbf{M}$ configuration and favours the cooperative switching of the ferromagnetic layers.

\subsubsection{Stability of parallel configuration $\mathbf{M}^*\uparrow\uparrow \mathbf{M}\uparrow\uparrow \mathbf{N}$}

Up to now we discussed only the structure of the equilibrium states. However, field-induced switching is associated with the stability of states with respect to thermal fluctuations. In the simple picture of a single superparamagnetic particle \cite{Brown1963,Peng2004}, magnetization switches once the magnetic field compensates the magnetic anisotropy and the frequency of the long-wave magnons is close to zero. To address the magnetic stability of the trilayer, we thus analyse the magnon spectra  around the equilibrium  parallel configuration $\mathbf{M}^*\uparrow\uparrow \mathbf{M}\uparrow\uparrow\mathbf{N}$, searching for nonlocalized magnon modes able to correlate thermal fluctuations in F and F*

{\bf{\it{Method of calculation}}}. The magnon spectra are calculated by solving the standard dynamic equations for magnetic vectors  $\mathbf{M}^*$, $\mathbf{M}$, and $\mathbf{N}$. For the ferromagnetic layers, we use the Landau-Lifshitz equations:
\begin{equation} \label{eq_LLG_FM}
\begin{split}
\dot{\mathbf{M}} &= -\gamma \mathbf{M} \times  \left(\mathbf{H}_ \mathbf{M}+A\Delta \mathbf{M}\right),\\
\dot{\mathbf{M}}^* &= -\gamma \mathbf{M}^* \times \left( \mathbf{H}_ {\mathbf{M}^*}+A\Delta \mathbf{M}^*\right),
\end{split}
\end{equation}
where $\gamma$ is the gyromagnetic ratio, the effective anisotropy field is $\mathbf{H}_ \mathbf{M}=-\partial w_\mathrm{F}/\partial \mathbf{M}$ ($\mathbf{H}_ {\mathbf{M}^*}=-\partial w_\mathrm{F*}/\partial {\mathbf{M}^*}$), and $w_\mathrm{F}$ ($w_\mathrm{F*}$) is the magnetic energy density introduced by Eq.~\eqref{eq_ferromagnetic-energy}.
 
For the N\'eel vector of the antiferromagnetic layer we use the standard AF-dynamic equation (see, e.g.\cite{Gomonay2014b}):
 \begin{equation}\label{eq_AF}
\mathbf{N}\times\left(\ddot{\mathbf{N}}-c^2\Delta \mathbf{N}+\gamma^2H_\mathrm{ex}\mathbf{H}_\mathbf{N}\right)=0,
\end{equation}
where $c=\gamma \sqrt{A_\mathrm{AF}H_\mathrm{ex}M_s}$ is the limiting magnon velocity in the antiferromagnet. We also introduce the effective anisotropy field of the antiferromagnetic layer as $\mathbf{H}_ \mathbf{N}=-\partial w_\mathrm{AF}/\partial \mathbf{N}$, where $w_\mathrm{AF}$ is given by Eq.~\eqref{eq_antiferromagnetic-energy}.

Equations \eqref{eq_LLG_FM} and \eqref{eq_AF} are complemented by the exchange boundary conditions, $\partial_x\mathbf{M}\mid_{x=L_\mathrm{F}}=0$ and $\partial_x\mathbf{M^*}\mid_{x=-L_\mathrm{F*}}=0$, at the free surfaces of the ferromagnets, and the following relations at the AF/F interface:
\begin{equation}\label{eq_boundary_conditions}
\begin{split}
 \left(A\partial_x\mathbf{M}+\frac{\partial w_\mathrm{int}}{\partial \mathbf{M}}\right)_{x=t_\mathrm{AF}/2}&=0,\\ \left(A_\mathrm{AF}\partial_x\mathbf{N}+\frac{\partial w_\mathrm{int}}{\partial \mathbf{M}}\right)_{x=t_\mathrm{AF}/2}&=0
\end{split}
\end{equation}
with similar relations at the F*/AF interface.

Magnons are introduced as small excitations $\delta \mathbf{M}^*$,  $\delta \mathbf{M}$, and $\delta \mathbf{N}$ over the equilibrium state with $\mathbf{M}_0^*\uparrow\uparrow\mathbf{M}_0\uparrow\uparrow\mathbf{N}_0$. Magnon spectra of a trilayer are calculated using approach suggested in Ref.\cite{Lifshitz1955}.

For standalone F (F*) and AF layers the eigen modes are obviously localised  within the corresponding layers. The eigen-modes are classified according to the wave-vector $k$, $k_*$ in the ferromagnetic layers and $q$ in the antiferromagnetic layer. The eigen-frequencies calculated according to Eqs.~\eqref{eq_LLG_FM} and \eqref{eq_AF} are given by the Kittel's formula for F and F*:
\begin{equation}\label{eq_Kittel} 
\begin{split}
\omega_F &= \gamma\sqrt{4\pi M(K_\mathrm{an}M+H)+AMk^2},\\ \omega_{F^*} &= \gamma\sqrt{4\pi M^*(K^*_\mathrm{an}M^*+H)+AM^*k_*^2},
\end{split}
\end{equation}
where we assume that $4\pi M\gg H, K_\mathrm{an}M$, and 
\begin{equation}\label{eq_frequency_AF}
\omega_\mathrm{AF}=\gamma \sqrt{H_\mathrm{ex}K_\mathrm{AF}M_s+c^2q^2}
\end{equation}
for the antiferromagnetic layer.

{\bf{\it{Analysis of results and discussion}}}. The magnon modes of the trilayer are hybridised due to the coupling at the interfaces so the wave-vectors in different layers are no longer independent. For a given frequency of an eigen-mode $\omega$, 
$q=\sqrt{\omega^2-\omega^2_\mathrm{AF}}/c$,  and 	
\begin{equation}\label{eq_dispersion_relation_FM1}
	k=\frac{\sqrt{\omega-\omega_F}}{\sqrt{\gamma AM}},\quad k_*=\frac{\sqrt{\omega-\omega_{F^*}}}{\sqrt{\gamma AM^*}}.
	\end{equation}

The eigenmode frequencies then calculated from
\begin{eqnarray}\label{eq_eigen_modes}
&&\left[(k-\lambda)(k_*+\lambda^*)(q-\bar{\lambda})(q+\bar{\lambda}^*)-\right.\\&&\left.-\frac{\lambda\bar{\lambda}\lambda^*\bar{\lambda}^*}{\tan(kL_{F})\tan(k_*L^*_{F})}\right]\frac{\sin qt_\mathrm{AF}}{qt_\mathrm{AF}}\nonumber\\&&+\left[\frac{\lambda\bar{\lambda}(k_*+\lambda^*)}{\tan(kL_{F})}-\frac{\lambda^*\bar{\lambda}^*(k-\lambda)}{\tan(k_*L^*_{F})}\right]\frac{\cos qt_\mathrm{AF}}{t_\mathrm{AF}}=0,\nonumber
\end{eqnarray}
with $q\Rightarrow i\kappa$ for $\omega<\omega_\mathrm{AF}$. 

In Eq.~\eqref{eq_eigen_modes} we introduced effective interaction constants: 
\begin{equation} \label{eq_notations4_interaction_constants}
\begin{split}
%	\lambda_j\equiv\frac{\xi H_{jb}M_j}{\alpha_\mathrm{Fj}},\quad \bar{\lambda}_j\equiv\frac{\xi H_{jb}M_j}{\alpha_\mathrm{AF}}
\lambda &\equiv \frac{\xi J_{b}M_s}{AM},\quad 
\lambda^*\equiv \frac{\xi J^*_{b}M_s}{AM^*},\\ \bar{\lambda} &\equiv \frac{\xi J_{b}M}{A_\mathrm{AF}M_s}\quad \bar{\lambda}^*\equiv \frac{\xi J^*_{b}M_*}{A_\mathrm{AF}M_s}.
\end{split}
\end{equation} 

$\lambda$ and $\lambda^*$ parametrize the momentum transferred by the ferromagnetic magnons to the antiferromagnetic layer in the limit of an infinitely hard antiferromagnet. As such, they define the contribution of the interface exchange coupling to the effective  anisotropy of the ferromagnetic layers. In a similar way, $\bar{\lambda}$ and $\bar{\lambda}^*$ quantify the momentum absorbed by the antiferromagnetic magnons.  

\begin{figure}%[htbp]
\begin{center}
\includegraphics[width=8.5 cm]{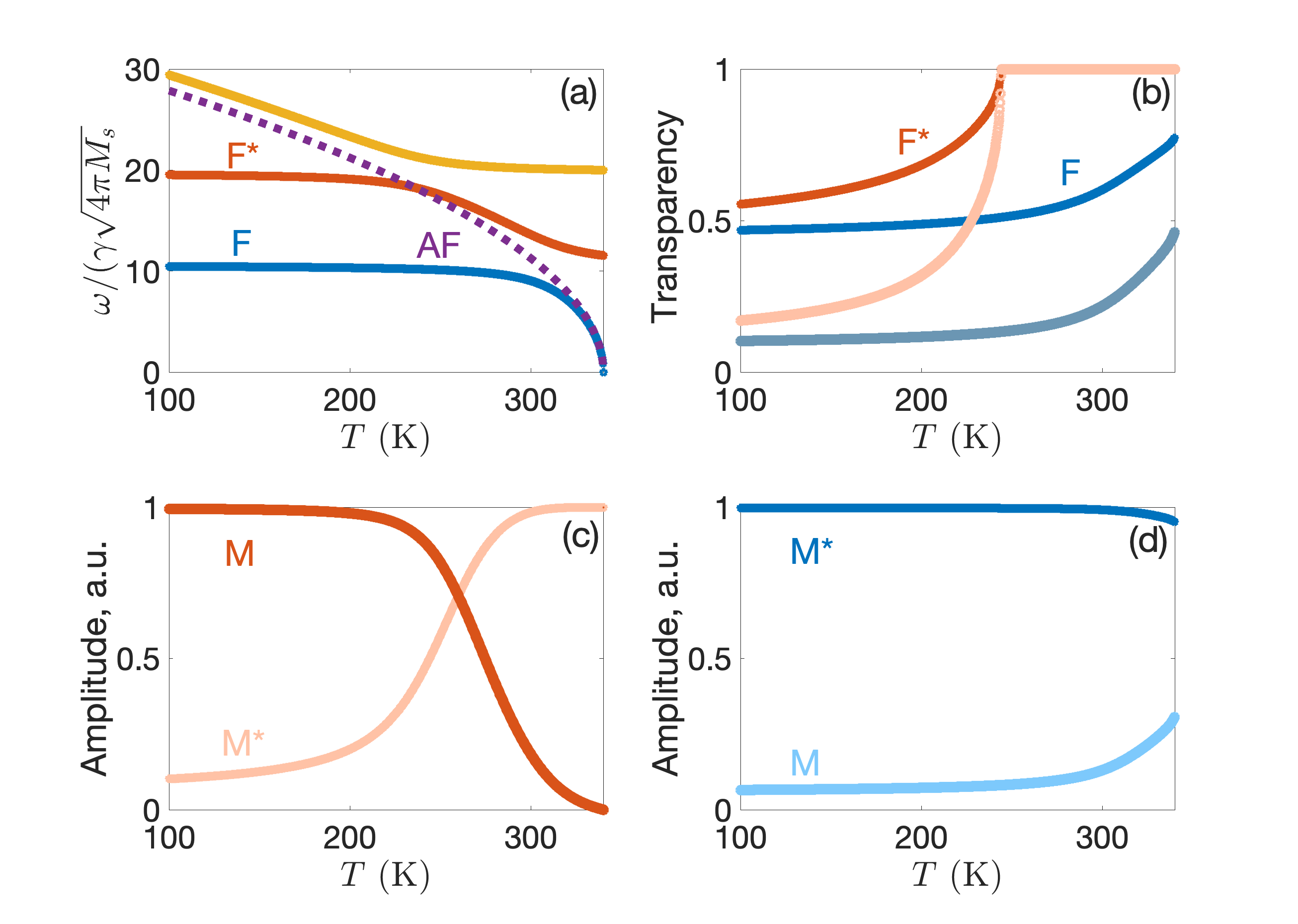}
\caption{Magnon spectra of F*/AF/F trilayer at different temperatures calculated for $t_\mathrm{AF}=6$~nm. (a) Hybridized magnon modes. Dash line shows temperature dependence of $\omega_\mathrm{AF}$ of isolated AF layer. (b) Transparency of AF layer calculated according to \eqref{eq_transparency} for quasi-F (blue) and quasi-F* (orange) modes for $t_\mathrm{AF}=$6~nm (bright colors) and $t_\mathrm{AF}=$12~nm (dim colors). Quasi-F mode is hybridized with evanacent AF magnons in all temperature range, hence $\mathcal{T}<1$. (c),(d) Relative amplitudes of $M$ and $M^*$ in quasi-F* (c) and quasi-F (d) modes. At low temperature, quasi-F* (quasi-F) mode is localised in F* (F) layer. At $T_1<T<T_2$, quasi-F* mode is delocalized, while quasi-F mode is still mostly localized within F layer.}
\label{fig_spectra}
\end{center}
\end{figure}

The magnon spectra of a F*/AF/F trilayer, given by Eq.~\eqref{eq_eigen_modes}, consist of three branches corresponding to two (quasi-)ferromagnetic and one quasi-antiferromagnetic modes~\footnote{~The magnon spectra of an antiferromagnet consist of two branches with different polarization. However, in the present model we consider only one branch, which hybridises with the ferromagnetic magnons at the interfaces.} The structure of the modes and the extent of their hybridization depend on the relation between the frequencies $\omega_\mathrm{AF}$, $\omega_\mathrm{F}$, and the interaction constants \eqref{eq_notations4_interaction_constants}, whose values are temperature dependent. Here we consider three different regimes relevant for the experimental observations detailed above assuming that $\omega_\mathrm{AF}(T)$ varies in a wide range of values due to the temperature dependence of the antiferromagnetic order parameter, $M_s\propto\sqrt{T_\mathrm{N}-T}$, where $T_\mathrm{N}$ is the N\'eel temperature (see Fig.~\ref{fig_spectra}). These regimes are delimited by temperature $T_1$ of crossing of the ferro- and antiferro-magnetic spectra, $\omega_\mathrm{AF}(T_1)=\omega_\mathrm{F*}$,  and temperature $T_2$, $\xi J^*_{b}M_s(T_2)=4\pi M\sqrt{K_\mathrm{an}}$, at which the exchange coupling at the interfaces becomes negligible. We have $T_1<T_2 \le T_\mathrm{N}$.  

\begin{figure}%[htbp]
\begin{center}
\includegraphics[width=8.5 cm]{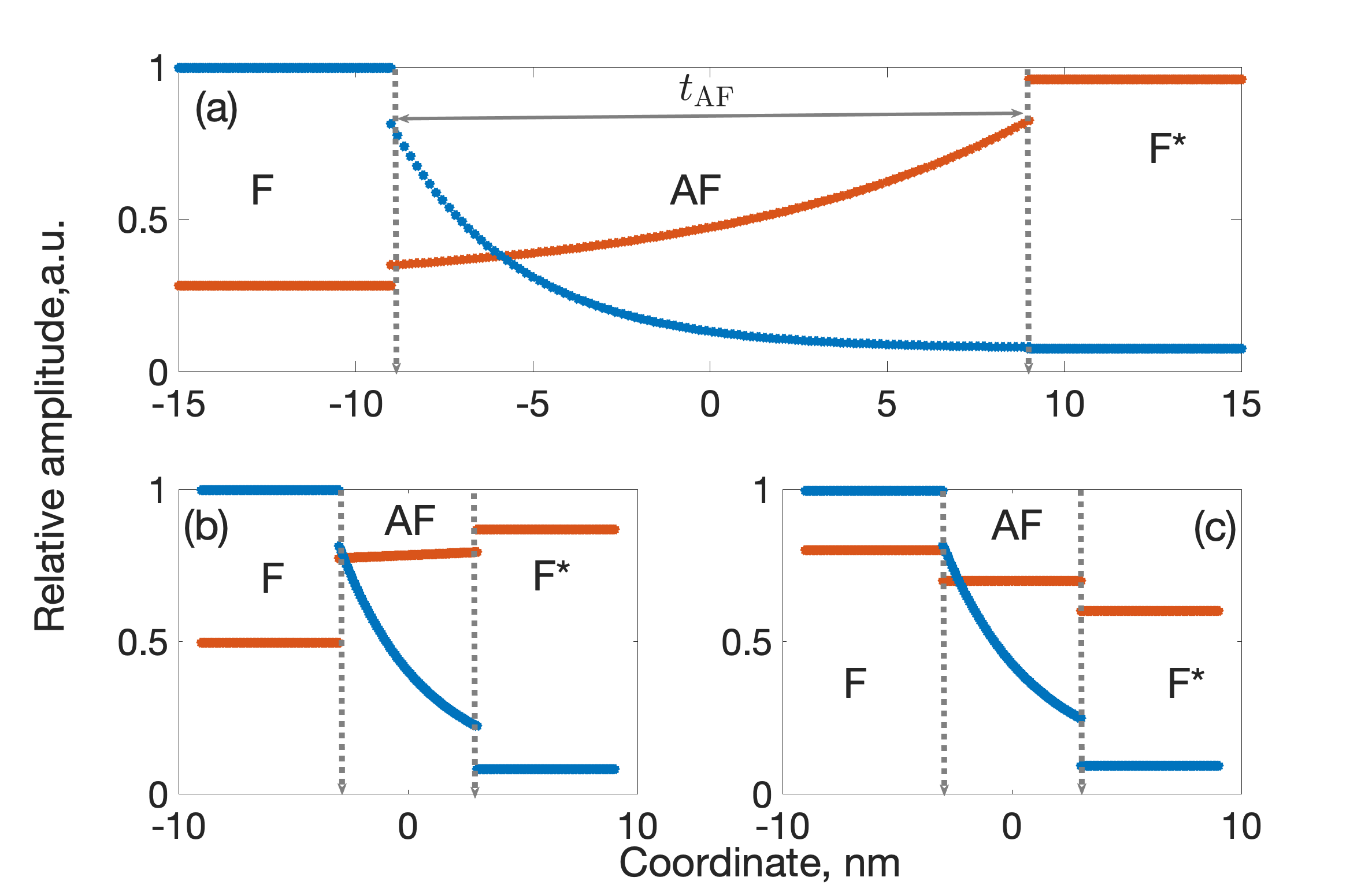}
\caption{Spatial distribution of amplitudes in quasi-F (blue lines) and quasi-F* (orange lines) magnon modes at different temperatures and $t_\mathrm{AF}$. (a) $T=220$~K$< T_1$, $t_\mathrm{AF}=12$~nm. Oscillations are localized within F and F* layers, amplitude of N\'eel vector exponentially decays. (b) $T_1\le T=245$~K, $t_\mathrm{AF}=6$~nm. (c) $T_1\le T=270$~K, $t_\mathrm{AF}=6$~nm. AF layer is almost transparent for quasi-F* magnons. Discontinuity at interfaces correspond to momentum exchange between $\mathbf{N}$ and $\mathbf{M}$ ($\mathbf{M^*}$) at interface.}
\label{fig_amplitude}
\end{center}
\end{figure}

At low temperature, $T<T_1$, the antiferromagnet is magnetically hard, $\omega_\mathrm{AF}\gg \omega_\mathrm{F}, \omega_\mathrm{F*}$, and hybridization of the modes is weak. The quasi-ferromagnetic modes, with $\omega\approx \omega_\mathrm{F*}$ and $\omega\approx \omega_\mathrm{F}$, which are responsible for the field-induced switching, are mainly localised within the ferromagnetic layers, with the ''tail'' of the evanescent antiferromagnetic mode decaying as $\propto \exp{\left[-(t_\mathrm{AF}/2+ x)\sqrt{\omega^2_\mathrm{AF}-\omega_\mathrm{F*}^2}/c\right]}$ or $\propto \exp{\left[-(t_\mathrm{AF}/2- x)\sqrt{\omega^2_\mathrm{AF}-\omega_\mathrm{F}^2}/c\right]}$, off the interface into F(F*) (see Fig.~\ref{fig_amplitude}). In this case, the dynamics of the ferromagnetic layers are fully decoupled.

In the intermediate temperature range, $T_1\le T\le T_2$, $\omega_\mathrm{F*}\ge\omega_\mathrm{AF}\propto \omega_\mathrm{F}$ and  hybridization of the modes is pronounced. The quasi-antiferromagnetic mode that results from hybridization of the F* and the propagating AF modes is fully delocalized between the F and F* layers. The quasi-F-mode is also delocalized  if $\omega_\mathrm{AF}\le\omega_\mathrm{F}$. In the opposite case, $\omega_\mathrm{AF}\ge\omega_\mathrm{F}$, the evanescent antiferromagnetic mode transmits spin excitations through the layer (see Fig.~\ref{fig_spectra}b), with the transparency coefficient given by
\begin{equation}\label{eq_transparency}
\mathcal{T}\propto\exp\left({-\sqrt{\omega^2_\mathrm{F}-\omega^2_\mathrm{AF}}t_\mathrm{AF}/c}\right).
\end{equation}
In this temperature range the dynamics of the ferromagnetic layers are fully coupled.

It should be noted that delocalization of the quasi-ferromagnetic modes hybridized with the evanescent antiferromagnetic mode can be controlled both by temperature and by thickness of the antiferromagnetic layer, as is illustrated in Figs.~\ref{fig_spectra},\ref{fig_amplitude}. For example, in thick films with $t_\mathrm{AF}\gg x_\mathrm{DW}$, transparency $\mathcal{T}$ is exponentially small in almost all of the temperature range below $T_\mathrm{N}$.

At high temperature, $T>T_2$, the AF layer is soft ($\omega_\mathrm{AF}\le \omega_\mathrm{F}<\omega_\mathrm{F*}$) and transparent to the F and F* magnons. In this limit, however, the effective coupling at the F/AF and AF/F* interfaces is significantly weakened by strong thermal fluctuations of the AF order and the resulting much reduced value of the N\'eel vector, $|\mathbf{N}|\rightarrow 0$. This behavior is also consistent with the small enery barrier between parallel and antiparallel states of ferromangetic magnetizations, since for this temperature range $\xi J_b M/E_\mathrm{DW}<1$ (see Fig.~\ref{fig_energy_barrier_DW}). As a result, correlations between the ferromagnetic layers vanish also at high temperature.

From the above theoretical analysis we conclude that a {\it{thin and intermediate}}  AF spacer can effectively mediate coupling between two ferromagnetic layers in a finite temperature range near $T_\mathrm{N}$, in which the AF is soft enough to transmit the F and F* magnons while thermal fluctuations are too weak to suppress the exchange interactions at the F/AF and AF/F* interfaces. As regards two more straightforward cases showing no transitions between the different thermo-magnetic regimes, {\it{ultrathin}} AF layers mediate strong F-F* coupling whereas {\it{thick}} AF layers fully decouple the outer F-layers, in the entire temperature range, even above $T_\mathrm{N}$. 

\section{Conclusions}

Interlayer coupling via AF spacers in multilayers Fe*/FeMn($t$)/Py ($t =$ 4-15~nm) is investigated. It is shown that the thickness of the FeMn spacer and ultimately its magnetic state play a decisive role in determining the behavior of the system as regards the interlayer exchange. The observed temperature and thickness dependence of the interlayer coupling can be explained in terms of the interplay between the intrinsic AF-ordering and the extrinsic finite-size and priximity effects.  

Particularly strong interlayer coupling with no exchange bias is observed for the structure with thin FeMn ($t <$ 5~nm) even at temperatures significantly exceeding the effective Néel temperature of the spacer (up to 450~K for $t =$ 6~nm, with, e.g., $T_\mathrm{N} \approx$ 320~K for $t =$ 6~nm). We explain this interlayer coupling as mainly governed by a strong magnetic proximity effect from the outer Fe* and Py. 

The structures with thicker FeMn ($t \geq$ 6~nm) offer the possibility of thermal gating of the interlayer exchange, which is significant only in the vicinity of the Néel point. Thicker spacers behave as expected of bulk antiferromagnets, offering strong exchange-bias at all temperatures below the Néel transition. 

The established three-region phase space for the  interlayer-coupling vs AF-thickness should broaden the understanding of the mechanisms at play in AF-based nanostructures and help develop novel devices for antiferromagnetic spintronics. 

\begin{acknowledgments}
Support from the Swedish Research Council (VR:2018-03526), the Swedish Strategic Research Council (SSF UKR22-0050), the Wenner-Gren Foundation (grant GFU2022-0011), the Olle Engkvist Foundation (2020:207-0460), the Volkswagen Foundation (Grant No. 97758), and the National Academy of Sciences of Ukraine (No. 0122U002260) are gratefully acknowledged. Y.L. acknowledges support from the Central European Research Infrastructure (CERIC) Consortium (Horizon 2020, project ACCELERATE, No. 731112). O.G. acknowledges support from the Alexander von Humboldt Foundation, the ERC Synergy Grant SC2 (No. 610115), and the Deutsche Forschungsgemeinschaft (DFG, German Research Foundation) - TRR 173 – 268565370 (project B12). 
\end{acknowledgments}

\bibliography{Refs}

\begin{thebibliography}{53}%
\makeatletter
\providecommand \@ifxundefined [1]{%
 \@ifx{#1\undefined}
}%
\providecommand \@ifnum [1]{%
 \ifnum #1\expandafter \@firstoftwo
 \else \expandafter \@secondoftwo
 \fi
}%
\providecommand \@ifx [1]{%
 \ifx #1\expandafter \@firstoftwo
 \else \expandafter \@secondoftwo
 \fi
}%
\providecommand \natexlab [1]{#1}%
\providecommand \enquote  [1]{``#1''}%
\providecommand \bibnamefont  [1]{#1}%
\providecommand \bibfnamefont [1]{#1}%
\providecommand \citenamefont [1]{#1}%
\providecommand \href@noop [0]{\@secondoftwo}%
\providecommand \href [0]{\begingroup \@sanitize@url \@href}%
\providecommand \@href[1]{\@@startlink{#1}\@@href}%
\providecommand \@@href[1]{\endgroup#1\@@endlink}%
\providecommand \@sanitize@url [0]{\catcode `\\12\catcode `\$12\catcode
  `\&12\catcode `\#12\catcode `\^12\catcode `\_12\catcode `\%12\relax}%
\providecommand \@@startlink[1]{}%
\providecommand \@@endlink[0]{}%
\providecommand \url  [0]{\begingroup\@sanitize@url \@url }%
\providecommand \@url [1]{\endgroup\@href {#1}{\urlprefix }}%
\providecommand \urlprefix  [0]{URL }%
\providecommand \Eprint [0]{\href }%
\providecommand \doibase [0]{http://dx.doi.org/}%
\providecommand \selectlanguage [0]{\@gobble}%
\providecommand \bibinfo  [0]{\@secondoftwo}%
\providecommand \bibfield  [0]{\@secondoftwo}%
\providecommand \translation [1]{[#1]}%
\providecommand \BibitemOpen [0]{}%
\providecommand \bibitemStop [0]{}%
\providecommand \bibitemNoStop [0]{.\EOS\space}%
\providecommand \EOS [0]{\spacefactor3000\relax}%
\providecommand \BibitemShut  [1]{\csname bibitem#1\endcsname}%
\let\auto@bib@innerbib\@empty
%</preamble>
\bibitem [{\citenamefont {Jungfleisch}\ \emph {et~al.}(2018)\citenamefont
  {Jungfleisch}, \citenamefont {Zhang},\ and\ \citenamefont
  {Hoffmann}}]{Jungfleisch2018}%
  \BibitemOpen
  \bibfield  {author} {\bibinfo {author} {\bibfnamefont {Matthias~B.}\
  \bibnamefont {Jungfleisch}}, \bibinfo {author} {\bibfnamefont {Wei}\
  \bibnamefont {Zhang}}, \ and\ \bibinfo {author} {\bibfnamefont {Axel}\
  \bibnamefont {Hoffmann}},\ }\bibfield  {title} {\enquote {\bibinfo {title}
  {Perspectives of antiferromagnetic spintronics},}\ }\href {\doibase
  10.1016/j.physleta.2018.01.008} {\bibfield  {journal} {\bibinfo  {journal}
  {Physics Letters A}\ }\textbf {\bibinfo {volume} {382}},\ \bibinfo {pages}
  {865--871} (\bibinfo {year} {2018})}\BibitemShut {NoStop}%
\bibitem [{\citenamefont {Wang}\ \emph {et~al.}(2014)\citenamefont {Wang},
  \citenamefont {Song}, \citenamefont {Wang}, \citenamefont {Miao},
  \citenamefont {Zeng},\ and\ \citenamefont {Pan}}]{Wang2014}%
  \BibitemOpen
  \bibfield  {author} {\bibinfo {author} {\bibfnamefont {Yuyan}\ \bibnamefont
  {Wang}}, \bibinfo {author} {\bibfnamefont {Cheng}\ \bibnamefont {Song}},
  \bibinfo {author} {\bibfnamefont {Guangyue}\ \bibnamefont {Wang}}, \bibinfo
  {author} {\bibfnamefont {Jinghui}\ \bibnamefont {Miao}}, \bibinfo {author}
  {\bibfnamefont {Fei}\ \bibnamefont {Zeng}}, \ and\ \bibinfo {author}
  {\bibfnamefont {Feng}\ \bibnamefont {Pan}},\ }\bibfield  {title} {\enquote
  {\bibinfo {title} {Anti-ferromagnet controlled tunneling
  magnetoresistance},}\ }\href {\doibase 10.1002/adfm.201401659} {\bibfield
  {journal} {\bibinfo  {journal} {Advanced Functional Materials}\ }\textbf
  {\bibinfo {volume} {24}},\ \bibinfo {pages} {6806--6810} (\bibinfo {year}
  {2014})}\BibitemShut {NoStop}%
\bibitem [{\citenamefont {Marti}\ \emph {et~al.}(2014)\citenamefont {Marti},
  \citenamefont {Fina}, \citenamefont {Frontera}, \citenamefont {Liu},
  \citenamefont {Wadley}, \citenamefont {He}, \citenamefont {Paull},
  \citenamefont {Clarkson}, \citenamefont {Kudrnovsk{\'{y}}}, \citenamefont
  {Turek}, \citenamefont {Kune{\v{s}}}, \citenamefont {Yi}, \citenamefont
  {Chu}, \citenamefont {Nelson}, \citenamefont {You}, \citenamefont {Arenholz},
  \citenamefont {Salahuddin}, \citenamefont {Fontcuberta}, \citenamefont
  {Jungwirth},\ and\ \citenamefont {Ramesh}}]{Marti2014}%
  \BibitemOpen
  \bibfield  {author} {\bibinfo {author} {\bibfnamefont {X.}~\bibnamefont
  {Marti}}, \bibinfo {author} {\bibfnamefont {I.}~\bibnamefont {Fina}},
  \bibinfo {author} {\bibfnamefont {C.}~\bibnamefont {Frontera}}, \bibinfo
  {author} {\bibfnamefont {Jian}\ \bibnamefont {Liu}}, \bibinfo {author}
  {\bibfnamefont {P.}~\bibnamefont {Wadley}}, \bibinfo {author} {\bibfnamefont
  {Q.}~\bibnamefont {He}}, \bibinfo {author} {\bibfnamefont {R.~J.}\
  \bibnamefont {Paull}}, \bibinfo {author} {\bibfnamefont {J.~D.}\ \bibnamefont
  {Clarkson}}, \bibinfo {author} {\bibfnamefont {J.}~\bibnamefont
  {Kudrnovsk{\'{y}}}}, \bibinfo {author} {\bibfnamefont {I.}~\bibnamefont
  {Turek}}, \bibinfo {author} {\bibfnamefont {J.}~\bibnamefont {Kune{\v{s}}}},
  \bibinfo {author} {\bibfnamefont {D.}~\bibnamefont {Yi}}, \bibinfo {author}
  {\bibfnamefont {J-H.}\ \bibnamefont {Chu}}, \bibinfo {author} {\bibfnamefont
  {C.~T.}\ \bibnamefont {Nelson}}, \bibinfo {author} {\bibfnamefont
  {L.}~\bibnamefont {You}}, \bibinfo {author} {\bibfnamefont {E.}~\bibnamefont
  {Arenholz}}, \bibinfo {author} {\bibfnamefont {S.}~\bibnamefont
  {Salahuddin}}, \bibinfo {author} {\bibfnamefont {J.}~\bibnamefont
  {Fontcuberta}}, \bibinfo {author} {\bibfnamefont {T.}~\bibnamefont
  {Jungwirth}}, \ and\ \bibinfo {author} {\bibfnamefont {R.}~\bibnamefont
  {Ramesh}},\ }\bibfield  {title} {\enquote {\bibinfo {title} {Room-temperature
  antiferromagnetic memory resistor},}\ }\href {\doibase 10.1038/nmat3861}
  {\bibfield  {journal} {\bibinfo  {journal} {Nature Materials}\ }\textbf
  {\bibinfo {volume} {13}},\ \bibinfo {pages} {367--374} (\bibinfo {year}
  {2014})}\BibitemShut {NoStop}%
\bibitem [{\citenamefont {Kriegner}\ \emph {et~al.}(2016)\citenamefont
  {Kriegner}, \citenamefont {V{\'{y}}born{\'{y}}}, \citenamefont
  {Olejn{\'{\i}}k}, \citenamefont {Reichlov{\'{a}}}, \citenamefont
  {Nov{\'{a}}k}, \citenamefont {Marti}, \citenamefont {Gazquez}, \citenamefont
  {Saidl}, \citenamefont {N{\v{e}}mec}, \citenamefont {Volobuev}, \citenamefont
  {Springholz}, \citenamefont {Hol{\'{y}}},\ and\ \citenamefont
  {Jungwirth}}]{Kriegner2016}%
  \BibitemOpen
  \bibfield  {author} {\bibinfo {author} {\bibfnamefont {D.}~\bibnamefont
  {Kriegner}}, \bibinfo {author} {\bibfnamefont {K.}~\bibnamefont
  {V{\'{y}}born{\'{y}}}}, \bibinfo {author} {\bibfnamefont {K.}~\bibnamefont
  {Olejn{\'{\i}}k}}, \bibinfo {author} {\bibfnamefont {H.}~\bibnamefont
  {Reichlov{\'{a}}}}, \bibinfo {author} {\bibfnamefont {V.}~\bibnamefont
  {Nov{\'{a}}k}}, \bibinfo {author} {\bibfnamefont {X.}~\bibnamefont {Marti}},
  \bibinfo {author} {\bibfnamefont {J.}~\bibnamefont {Gazquez}}, \bibinfo
  {author} {\bibfnamefont {V.}~\bibnamefont {Saidl}}, \bibinfo {author}
  {\bibfnamefont {P.}~\bibnamefont {N{\v{e}}mec}}, \bibinfo {author}
  {\bibfnamefont {V.~V.}\ \bibnamefont {Volobuev}}, \bibinfo {author}
  {\bibfnamefont {G.}~\bibnamefont {Springholz}}, \bibinfo {author}
  {\bibfnamefont {V.}~\bibnamefont {Hol{\'{y}}}}, \ and\ \bibinfo {author}
  {\bibfnamefont {T.}~\bibnamefont {Jungwirth}},\ }\bibfield  {title} {\enquote
  {\bibinfo {title} {Multiple-stable anisotropic magnetoresistance memory in
  antiferromagnetic {MnTe}},}\ }\href {\doibase 10.1038/ncomms11623} {\bibfield
   {journal} {\bibinfo  {journal} {Nature Communications}\ }\textbf {\bibinfo
  {volume} {7}},\ \bibinfo {pages} {11623} (\bibinfo {year}
  {2016})}\BibitemShut {NoStop}%
\bibitem [{\citenamefont {Bodnar}\ \emph {et~al.}(2020)\citenamefont {Bodnar},
  \citenamefont {Skourski}, \citenamefont {Gomonay}, \citenamefont {Sinova},
  \citenamefont {Kläui},\ and\ \citenamefont {Jourdan}}]{Bodnar2020}%
  \BibitemOpen
  \bibfield  {author} {\bibinfo {author} {\bibfnamefont {S.~Yu.}\ \bibnamefont
  {Bodnar}}, \bibinfo {author} {\bibfnamefont {Y.}~\bibnamefont {Skourski}},
  \bibinfo {author} {\bibfnamefont {O.}~\bibnamefont {Gomonay}}, \bibinfo
  {author} {\bibfnamefont {J.}~\bibnamefont {Sinova}}, \bibinfo {author}
  {\bibfnamefont {M.}~\bibnamefont {Kläui}}, \ and\ \bibinfo {author}
  {\bibfnamefont {M.}~\bibnamefont {Jourdan}},\ }\bibfield  {title} {\enquote
  {\bibinfo {title} {Magnetoresistance effects in the metallic antiferromagnet
  {Mn$_2$Au}},}\ }\href {\doibase 10.1103/physrevapplied.14.014004} {\bibfield
  {journal} {\bibinfo  {journal} {Physical Review Applied}\ }\textbf {\bibinfo
  {volume} {14}},\ \bibinfo {pages} {014004} (\bibinfo {year}
  {2020})}\BibitemShut {NoStop}%
\bibitem [{\citenamefont {Shick}\ \emph {et~al.}(2010)\citenamefont {Shick},
  \citenamefont {Khmelevskyi}, \citenamefont {Mryasov}, \citenamefont
  {Wunderlich},\ and\ \citenamefont {Jungwirth}}]{Shick2010}%
  \BibitemOpen
  \bibfield  {author} {\bibinfo {author} {\bibfnamefont {A.~B.}\ \bibnamefont
  {Shick}}, \bibinfo {author} {\bibfnamefont {S.}~\bibnamefont {Khmelevskyi}},
  \bibinfo {author} {\bibfnamefont {O.~N.}\ \bibnamefont {Mryasov}}, \bibinfo
  {author} {\bibfnamefont {J.}~\bibnamefont {Wunderlich}}, \ and\ \bibinfo
  {author} {\bibfnamefont {T.}~\bibnamefont {Jungwirth}},\ }\bibfield  {title}
  {\enquote {\bibinfo {title} {Spin-orbit coupling induced anisotropy effects
  in bimetallic antiferromagnets: A route towards antiferromagnetic
  spintronics},}\ }\href {\doibase 10.1103/physrevb.81.212409} {\bibfield
  {journal} {\bibinfo  {journal} {Physical Review B}\ }\textbf {\bibinfo
  {volume} {81}},\ \bibinfo {pages} {212409} (\bibinfo {year}
  {2010})}\BibitemShut {NoStop}%
\bibitem [{\citenamefont {Wang}\ \emph {et~al.}(2012)\citenamefont {Wang},
  \citenamefont {Song}, \citenamefont {Cui}, \citenamefont {Wang},
  \citenamefont {Zeng},\ and\ \citenamefont {Pan}}]{Wang2012}%
  \BibitemOpen
  \bibfield  {author} {\bibinfo {author} {\bibfnamefont {Y.~Y.}\ \bibnamefont
  {Wang}}, \bibinfo {author} {\bibfnamefont {C.}~\bibnamefont {Song}}, \bibinfo
  {author} {\bibfnamefont {B.}~\bibnamefont {Cui}}, \bibinfo {author}
  {\bibfnamefont {G.~Y.}\ \bibnamefont {Wang}}, \bibinfo {author}
  {\bibfnamefont {F.}~\bibnamefont {Zeng}}, \ and\ \bibinfo {author}
  {\bibfnamefont {F.}~\bibnamefont {Pan}},\ }\bibfield  {title} {\enquote
  {\bibinfo {title} {Room-temperature perpendicular exchange coupling and
  tunneling anisotropic magnetoresistance in an antiferromagnet-based tunnel
  junction},}\ }\href {\doibase 10.1103/physrevlett.109.137201} {\bibfield
  {journal} {\bibinfo  {journal} {Physical Review Letters}\ }\textbf {\bibinfo
  {volume} {109}},\ \bibinfo {pages} {137201} (\bibinfo {year}
  {2012})}\BibitemShut {NoStop}%
\bibitem [{\citenamefont {Park}\ \emph {et~al.}(2011)\citenamefont {Park},
  \citenamefont {Wunderlich}, \citenamefont {Mart{\'{\i}}}, \citenamefont
  {Hol{\'{y}}}, \citenamefont {Kurosaki}, \citenamefont {Yamada}, \citenamefont
  {Yamamoto}, \citenamefont {Nishide}, \citenamefont {Hayakawa}, \citenamefont
  {Takahashi}, \citenamefont {Shick},\ and\ \citenamefont
  {Jungwirth}}]{Park2011}%
  \BibitemOpen
  \bibfield  {author} {\bibinfo {author} {\bibfnamefont {B.~G.}\ \bibnamefont
  {Park}}, \bibinfo {author} {\bibfnamefont {J.}~\bibnamefont {Wunderlich}},
  \bibinfo {author} {\bibfnamefont {X.}~\bibnamefont {Mart{\'{\i}}}}, \bibinfo
  {author} {\bibfnamefont {V.}~\bibnamefont {Hol{\'{y}}}}, \bibinfo {author}
  {\bibfnamefont {Y.}~\bibnamefont {Kurosaki}}, \bibinfo {author}
  {\bibfnamefont {M.}~\bibnamefont {Yamada}}, \bibinfo {author} {\bibfnamefont
  {H.}~\bibnamefont {Yamamoto}}, \bibinfo {author} {\bibfnamefont
  {A.}~\bibnamefont {Nishide}}, \bibinfo {author} {\bibfnamefont
  {J.}~\bibnamefont {Hayakawa}}, \bibinfo {author} {\bibfnamefont
  {H.}~\bibnamefont {Takahashi}}, \bibinfo {author} {\bibfnamefont {A.~B.}\
  \bibnamefont {Shick}}, \ and\ \bibinfo {author} {\bibfnamefont
  {T.}~\bibnamefont {Jungwirth}},\ }\bibfield  {title} {\enquote {\bibinfo
  {title} {A spin-valve-like magnetoresistance of an antiferromagnet-based
  tunnel junction},}\ }\href {\doibase 10.1038/nmat2983} {\bibfield  {journal}
  {\bibinfo  {journal} {Nature Materials}\ }\textbf {\bibinfo {volume} {10}},\
  \bibinfo {pages} {347--351} (\bibinfo {year} {2011})}\BibitemShut {NoStop}%
\bibitem [{\citenamefont {Wu}\ \emph {et~al.}(2016)\citenamefont {Wu},
  \citenamefont {Zhang}, \citenamefont {KC}, \citenamefont {Borisov},
  \citenamefont {Pearson}, \citenamefont {Jiang}, \citenamefont {Lederman},
  \citenamefont {Hoffmann},\ and\ \citenamefont {Bhattacharya}}]{Wu2016}%
  \BibitemOpen
  \bibfield  {author} {\bibinfo {author} {\bibfnamefont {Stephen~M.}\
  \bibnamefont {Wu}}, \bibinfo {author} {\bibfnamefont {Wei}\ \bibnamefont
  {Zhang}}, \bibinfo {author} {\bibfnamefont {Amit}\ \bibnamefont {KC}},
  \bibinfo {author} {\bibfnamefont {Pavel}\ \bibnamefont {Borisov}}, \bibinfo
  {author} {\bibfnamefont {John~E.}\ \bibnamefont {Pearson}}, \bibinfo {author}
  {\bibfnamefont {J.~Samuel}\ \bibnamefont {Jiang}}, \bibinfo {author}
  {\bibfnamefont {David}\ \bibnamefont {Lederman}}, \bibinfo {author}
  {\bibfnamefont {Axel}\ \bibnamefont {Hoffmann}}, \ and\ \bibinfo {author}
  {\bibfnamefont {Anand}\ \bibnamefont {Bhattacharya}},\ }\bibfield  {title}
  {\enquote {\bibinfo {title} {Antiferromagnetic spin {Seebeck} effect},}\
  }\href {\doibase 10.1103/physrevlett.116.097204} {\bibfield  {journal}
  {\bibinfo  {journal} {Physical Review Letters}\ }\textbf {\bibinfo {volume}
  {116}},\ \bibinfo {pages} {097204} (\bibinfo {year} {2016})}\BibitemShut
  {NoStop}%
\bibitem [{\citenamefont {Rezende}\ \emph {et~al.}(2016)\citenamefont
  {Rezende}, \citenamefont {Rodr{\'{\i}}guez-Su{\'{a}}rez},\ and\ \citenamefont
  {Azevedo}}]{Rezende2016}%
  \BibitemOpen
  \bibfield  {author} {\bibinfo {author} {\bibfnamefont {S.~M.}\ \bibnamefont
  {Rezende}}, \bibinfo {author} {\bibfnamefont {R.~L.}\ \bibnamefont
  {Rodr{\'{\i}}guez-Su{\'{a}}rez}}, \ and\ \bibinfo {author} {\bibfnamefont
  {A.}~\bibnamefont {Azevedo}},\ }\bibfield  {title} {\enquote {\bibinfo
  {title} {Theory of the spin seebeck effect in antiferromagnets},}\ }\href
  {\doibase 10.1103/physrevb.93.014425} {\bibfield  {journal} {\bibinfo
  {journal} {Physical Review B}\ }\textbf {\bibinfo {volume} {93}},\ \bibinfo
  {pages} {014425} (\bibinfo {year} {2016})}\BibitemShut {NoStop}%
\bibitem [{\citenamefont {Mendes}\ \emph {et~al.}(2014)\citenamefont {Mendes},
  \citenamefont {Cunha}, \citenamefont {Santos}, \citenamefont {Ribeiro},
  \citenamefont {Machado}, \citenamefont {Rodr{\'{\i}}guez-Su{\'{a}}rez},
  \citenamefont {Azevedo},\ and\ \citenamefont {Rezende}}]{Mendes2014}%
  \BibitemOpen
  \bibfield  {author} {\bibinfo {author} {\bibfnamefont {J.~B.~S.}\
  \bibnamefont {Mendes}}, \bibinfo {author} {\bibfnamefont {R.~O.}\
  \bibnamefont {Cunha}}, \bibinfo {author} {\bibfnamefont {O.~Alves}\
  \bibnamefont {Santos}}, \bibinfo {author} {\bibfnamefont {P.~R.~T.}\
  \bibnamefont {Ribeiro}}, \bibinfo {author} {\bibfnamefont {F.~L.~A.}\
  \bibnamefont {Machado}}, \bibinfo {author} {\bibfnamefont {R.~L.}\
  \bibnamefont {Rodr{\'{\i}}guez-Su{\'{a}}rez}}, \bibinfo {author}
  {\bibfnamefont {A.}~\bibnamefont {Azevedo}}, \ and\ \bibinfo {author}
  {\bibfnamefont {S.~M.}\ \bibnamefont {Rezende}},\ }\bibfield  {title}
  {\enquote {\bibinfo {title} {Large inverse spin hall effect in the
  antiferromagnetic metal {Ir$_{20}$Mn$_{80}$}},}\ }\href {\doibase
  10.1103/physrevb.89.140406} {\bibfield  {journal} {\bibinfo  {journal}
  {Physical Review B}\ }\textbf {\bibinfo {volume} {89}},\ \bibinfo {pages}
  {140406} (\bibinfo {year} {2014})}\BibitemShut {NoStop}%
\bibitem [{\citenamefont {Qu}\ \emph {et~al.}(2015)\citenamefont {Qu},
  \citenamefont {Huang},\ and\ \citenamefont {Chien}}]{Qu2015}%
  \BibitemOpen
  \bibfield  {author} {\bibinfo {author} {\bibfnamefont {D.}~\bibnamefont
  {Qu}}, \bibinfo {author} {\bibfnamefont {S.~Y.}\ \bibnamefont {Huang}}, \
  and\ \bibinfo {author} {\bibfnamefont {C.~L.}\ \bibnamefont {Chien}},\
  }\bibfield  {title} {\enquote {\bibinfo {title} {Inverse spin {Hall} effect
  in {Cr}: Independence of antiferromagnetic ordering},}\ }\href {\doibase
  10.1103/physrevb.92.020418} {\bibfield  {journal} {\bibinfo  {journal}
  {Physical Review B}\ }\textbf {\bibinfo {volume} {92}},\ \bibinfo {pages}
  {020418} (\bibinfo {year} {2015})}\BibitemShut {NoStop}%
\bibitem [{\citenamefont {Reichlov{\'{a}}}\ \emph {et~al.}(2015)\citenamefont
  {Reichlov{\'{a}}}, \citenamefont {Kriegner}, \citenamefont {Hol{\'{y}}},
  \citenamefont {Olejn{\'{\i}}k}, \citenamefont {Nov{\'{a}}k}, \citenamefont
  {Yamada}, \citenamefont {Miura}, \citenamefont {Ogawa}, \citenamefont
  {Takahashi}, \citenamefont {Jungwirth},\ and\ \citenamefont
  {Wunderlich}}]{Reichlova2015}%
  \BibitemOpen
  \bibfield  {author} {\bibinfo {author} {\bibfnamefont {H.}~\bibnamefont
  {Reichlov{\'{a}}}}, \bibinfo {author} {\bibfnamefont {D.}~\bibnamefont
  {Kriegner}}, \bibinfo {author} {\bibfnamefont {V.}~\bibnamefont
  {Hol{\'{y}}}}, \bibinfo {author} {\bibfnamefont {K.}~\bibnamefont
  {Olejn{\'{\i}}k}}, \bibinfo {author} {\bibfnamefont {V.}~\bibnamefont
  {Nov{\'{a}}k}}, \bibinfo {author} {\bibfnamefont {M.}~\bibnamefont {Yamada}},
  \bibinfo {author} {\bibfnamefont {K.}~\bibnamefont {Miura}}, \bibinfo
  {author} {\bibfnamefont {S.}~\bibnamefont {Ogawa}}, \bibinfo {author}
  {\bibfnamefont {H.}~\bibnamefont {Takahashi}}, \bibinfo {author}
  {\bibfnamefont {T.}~\bibnamefont {Jungwirth}}, \ and\ \bibinfo {author}
  {\bibfnamefont {J.}~\bibnamefont {Wunderlich}},\ }\bibfield  {title}
  {\enquote {\bibinfo {title} {Current-induced torques in structures with
  ultrathin {IrMn} antiferromagnets},}\ }\href {\doibase
  10.1103/physrevb.92.165424} {\bibfield  {journal} {\bibinfo  {journal}
  {Physical Review B}\ }\textbf {\bibinfo {volume} {92}},\ \bibinfo {pages}
  {165424} (\bibinfo {year} {2015})}\BibitemShut {NoStop}%
\bibitem [{\citenamefont {Zhang}\ \emph {et~al.}(2014)\citenamefont {Zhang},
  \citenamefont {Jungfleisch}, \citenamefont {Jiang}, \citenamefont {Pearson},
  \citenamefont {Hoffmann}, \citenamefont {Freimuth},\ and\ \citenamefont
  {Mokrousov}}]{Zhang2014}%
  \BibitemOpen
  \bibfield  {author} {\bibinfo {author} {\bibfnamefont {Wei}\ \bibnamefont
  {Zhang}}, \bibinfo {author} {\bibfnamefont {Matthias~B.}\ \bibnamefont
  {Jungfleisch}}, \bibinfo {author} {\bibfnamefont {Wanjun}\ \bibnamefont
  {Jiang}}, \bibinfo {author} {\bibfnamefont {John~E.}\ \bibnamefont
  {Pearson}}, \bibinfo {author} {\bibfnamefont {Axel}\ \bibnamefont
  {Hoffmann}}, \bibinfo {author} {\bibfnamefont {Frank}\ \bibnamefont
  {Freimuth}}, \ and\ \bibinfo {author} {\bibfnamefont {Yuriy}\ \bibnamefont
  {Mokrousov}},\ }\bibfield  {title} {\enquote {\bibinfo {title} {Spin {Hall}
  effects in metallic antiferromagnets},}\ }\href {\doibase
  10.1103/physrevlett.113.196602} {\bibfield  {journal} {\bibinfo  {journal}
  {Physical Review Letters}\ }\textbf {\bibinfo {volume} {113}},\ \bibinfo
  {pages} {196602} (\bibinfo {year} {2014})}\BibitemShut {NoStop}%
\bibitem [{\citenamefont {Nogu{\'{e}}s}\ and\ \citenamefont
  {Schuller}(1999)}]{Nogues1999}%
  \BibitemOpen
  \bibfield  {author} {\bibinfo {author} {\bibfnamefont {J}~\bibnamefont
  {Nogu{\'{e}}s}}\ and\ \bibinfo {author} {\bibfnamefont {Ivan~K}\ \bibnamefont
  {Schuller}},\ }\bibfield  {title} {\enquote {\bibinfo {title} {Exchange
  bias},}\ }\href {\doibase 10.1016/s0304-8853(98)00266-2} {\bibfield
  {journal} {\bibinfo  {journal} {Journal of Magnetism and Magnetic Materials}\
  }\textbf {\bibinfo {volume} {192}},\ \bibinfo {pages} {203--232} (\bibinfo
  {year} {1999})}\BibitemShut {NoStop}%
\bibitem [{\citenamefont {Wadley}\ \emph {et~al.}(2016)\citenamefont {Wadley},
  \citenamefont {Howells}, \citenamefont {elezny}, \citenamefont {Andrews},
  \citenamefont {Hills}, \citenamefont {Campion}, \citenamefont {Novak},
  \citenamefont {Olejnik}, \citenamefont {Maccherozzi}, \citenamefont {Dhesi},
  \citenamefont {Martin}, \citenamefont {Wagner}, \citenamefont {Wunderlich},
  \citenamefont {Freimuth}, \citenamefont {Mokrousov}, \citenamefont {Kune},
  \citenamefont {Chauhan}, \citenamefont {Grzybowski}, \citenamefont
  {Rushforth}, \citenamefont {Edmonds}, \citenamefont {Gallagher},\ and\
  \citenamefont {Jungwirth}}]{Wadley2016}%
  \BibitemOpen
  \bibfield  {author} {\bibinfo {author} {\bibfnamefont {P.}~\bibnamefont
  {Wadley}}, \bibinfo {author} {\bibfnamefont {B.}~\bibnamefont {Howells}},
  \bibinfo {author} {\bibfnamefont {J.}~\bibnamefont {elezny}}, \bibinfo
  {author} {\bibfnamefont {C.}~\bibnamefont {Andrews}}, \bibinfo {author}
  {\bibfnamefont {V.}~\bibnamefont {Hills}}, \bibinfo {author} {\bibfnamefont
  {R.~P.}\ \bibnamefont {Campion}}, \bibinfo {author} {\bibfnamefont
  {V.}~\bibnamefont {Novak}}, \bibinfo {author} {\bibfnamefont
  {K.}~\bibnamefont {Olejnik}}, \bibinfo {author} {\bibfnamefont
  {F.}~\bibnamefont {Maccherozzi}}, \bibinfo {author} {\bibfnamefont {S.~S.}\
  \bibnamefont {Dhesi}}, \bibinfo {author} {\bibfnamefont {S.~Y.}\ \bibnamefont
  {Martin}}, \bibinfo {author} {\bibfnamefont {T.}~\bibnamefont {Wagner}},
  \bibinfo {author} {\bibfnamefont {J.}~\bibnamefont {Wunderlich}}, \bibinfo
  {author} {\bibfnamefont {F.}~\bibnamefont {Freimuth}}, \bibinfo {author}
  {\bibfnamefont {Y.}~\bibnamefont {Mokrousov}}, \bibinfo {author}
  {\bibfnamefont {J.}~\bibnamefont {Kune}}, \bibinfo {author} {\bibfnamefont
  {J.~S.}\ \bibnamefont {Chauhan}}, \bibinfo {author} {\bibfnamefont {M.~J.}\
  \bibnamefont {Grzybowski}}, \bibinfo {author} {\bibfnamefont {A.~W.}\
  \bibnamefont {Rushforth}}, \bibinfo {author} {\bibfnamefont {K.~W.}\
  \bibnamefont {Edmonds}}, \bibinfo {author} {\bibfnamefont {B.~L.}\
  \bibnamefont {Gallagher}}, \ and\ \bibinfo {author} {\bibfnamefont
  {T.}~\bibnamefont {Jungwirth}},\ }\bibfield  {title} {\enquote {\bibinfo
  {title} {Electrical switching of an antiferromagnet},}\ }\href {\doibase
  10.1126/science.aab1031} {\bibfield  {journal} {\bibinfo  {journal}
  {Science}\ }\textbf {\bibinfo {volume} {351}},\ \bibinfo {pages} {587--590}
  (\bibinfo {year} {2016})}\BibitemShut {NoStop}%
\bibitem [{\citenamefont {Jungwirth}\ \emph {et~al.}(2016)\citenamefont
  {Jungwirth}, \citenamefont {Marti}, \citenamefont {Wadley},\ and\
  \citenamefont {Wunderlich}}]{Jungwirth2016}%
  \BibitemOpen
  \bibfield  {author} {\bibinfo {author} {\bibfnamefont {T.}~\bibnamefont
  {Jungwirth}}, \bibinfo {author} {\bibfnamefont {X.}~\bibnamefont {Marti}},
  \bibinfo {author} {\bibfnamefont {P.}~\bibnamefont {Wadley}}, \ and\ \bibinfo
  {author} {\bibfnamefont {J.}~\bibnamefont {Wunderlich}},\ }\bibfield  {title}
  {\enquote {\bibinfo {title} {Antiferromagnetic spintronics},}\ }\href
  {\doibase 10.1038/nnano.2016.18} {\bibfield  {journal} {\bibinfo  {journal}
  {Nature Nanotechnology}\ }\textbf {\bibinfo {volume} {11}},\ \bibinfo {pages}
  {231--241} (\bibinfo {year} {2016})}\BibitemShut {NoStop}%
\bibitem [{\citenamefont {Jungwirth}\ \emph {et~al.}(2018)\citenamefont
  {Jungwirth}, \citenamefont {Sinova}, \citenamefont {Manchon}, \citenamefont
  {Marti}, \citenamefont {Wunderlich},\ and\ \citenamefont
  {Felser}}]{Jungwirth2018}%
  \BibitemOpen
  \bibfield  {author} {\bibinfo {author} {\bibfnamefont {T.}~\bibnamefont
  {Jungwirth}}, \bibinfo {author} {\bibfnamefont {J.}~\bibnamefont {Sinova}},
  \bibinfo {author} {\bibfnamefont {A.}~\bibnamefont {Manchon}}, \bibinfo
  {author} {\bibfnamefont {X.}~\bibnamefont {Marti}}, \bibinfo {author}
  {\bibfnamefont {J.}~\bibnamefont {Wunderlich}}, \ and\ \bibinfo {author}
  {\bibfnamefont {C.}~\bibnamefont {Felser}},\ }\bibfield  {title} {\enquote
  {\bibinfo {title} {The multiple directions of antiferromagnetic
  spintronics},}\ }\href {\doibase 10.1038/s41567-018-0063-6} {\bibfield
  {journal} {\bibinfo  {journal} {Nature Physics}\ }\textbf {\bibinfo {volume}
  {14}},\ \bibinfo {pages} {200--203} (\bibinfo {year} {2018})}\BibitemShut
  {NoStop}%
\bibitem [{\citenamefont {Zhou}\ \emph {et~al.}(2020)\citenamefont {Zhou},
  \citenamefont {Liu},\ and\ \citenamefont {Wang}}]{Zhou2020}%
  \BibitemOpen
  \bibfield  {author} {\bibinfo {author} {\bibfnamefont {Z.~P.}\ \bibnamefont
  {Zhou}}, \bibinfo {author} {\bibfnamefont {X.~H.}\ \bibnamefont {Liu}}, \
  and\ \bibinfo {author} {\bibfnamefont {K.~Y.}\ \bibnamefont {Wang}},\
  }\bibfield  {title} {\enquote {\bibinfo {title} {Controlling vertical
  magnetization shift by spin{\textendash}orbit torque in
  ferromagnetic/antiferromagnetic/ferromagnetic heterostructure},}\ }\href
  {\doibase 10.1063/1.5139590} {\bibfield  {journal} {\bibinfo  {journal}
  {Applied Physics Letters}\ }\textbf {\bibinfo {volume} {116}},\ \bibinfo
  {pages} {062403} (\bibinfo {year} {2020})}\BibitemShut {NoStop}%
\bibitem [{\citenamefont {Stamps}(2000)}]{Stamps2000}%
  \BibitemOpen
  \bibfield  {author} {\bibinfo {author} {\bibfnamefont {R~L}\ \bibnamefont
  {Stamps}},\ }\bibfield  {title} {\enquote {\bibinfo {title} {Mechanisms for
  exchange bias},}\ }\href {\doibase 10.1088/0022-3727/33/23/201} {\bibfield
  {journal} {\bibinfo  {journal} {Journal of Physics D: Applied Physics}\
  }\textbf {\bibinfo {volume} {33}},\ \bibinfo {pages} {R247--R268} (\bibinfo
  {year} {2000})}\BibitemShut {NoStop}%
\bibitem [{\citenamefont {Kiwi}(2001)}]{Kiwi2001}%
  \BibitemOpen
  \bibfield  {author} {\bibinfo {author} {\bibfnamefont {Miguel}\ \bibnamefont
  {Kiwi}},\ }\bibfield  {title} {\enquote {\bibinfo {title} {Exchange bias
  theory},}\ }\href {\doibase 10.1016/s0304-8853(01)00421-8} {\bibfield
  {journal} {\bibinfo  {journal} {Journal of Magnetism and Magnetic Materials}\
  }\textbf {\bibinfo {volume} {234}},\ \bibinfo {pages} {584--595} (\bibinfo
  {year} {2001})}\BibitemShut {NoStop}%
\bibitem [{\citenamefont {Bobo}\ \emph {et~al.}(2004)\citenamefont {Bobo},
  \citenamefont {Gabillet},\ and\ \citenamefont {Bibes}}]{Bobo2004}%
  \BibitemOpen
  \bibfield  {author} {\bibinfo {author} {\bibfnamefont {J~F}\ \bibnamefont
  {Bobo}}, \bibinfo {author} {\bibfnamefont {L}~\bibnamefont {Gabillet}}, \
  and\ \bibinfo {author} {\bibfnamefont {M}~\bibnamefont {Bibes}},\ }\bibfield
  {title} {\enquote {\bibinfo {title} {Recent advances in nanomagnetism and
  spin electronics},}\ }\href {\doibase 10.1088/0953-8984/16/5/008} {\bibfield
  {journal} {\bibinfo  {journal} {Journal of Physics: Condensed Matter}\
  }\textbf {\bibinfo {volume} {16}},\ \bibinfo {pages} {S471--S496} (\bibinfo
  {year} {2004})}\BibitemShut {NoStop}%
\bibitem [{\citenamefont {Berkowitz}\ and\ \citenamefont
  {Takano}(1999)}]{Berkowitz1999}%
  \BibitemOpen
  \bibfield  {author} {\bibinfo {author} {\bibfnamefont {A.E.}\ \bibnamefont
  {Berkowitz}}\ and\ \bibinfo {author} {\bibfnamefont {Kentaro}\ \bibnamefont
  {Takano}},\ }\bibfield  {title} {\enquote {\bibinfo {title} {Exchange
  anisotropy {\textemdash} a review},}\ }\href {\doibase
  10.1016/s0304-8853(99)00453-9} {\bibfield  {journal} {\bibinfo  {journal}
  {Journal of Magnetism and Magnetic Materials}\ }\textbf {\bibinfo {volume}
  {200}},\ \bibinfo {pages} {552--570} (\bibinfo {year} {1999})}\BibitemShut
  {NoStop}%
\bibitem [{\citenamefont {Mohanty}\ \emph {et~al.}(2013)\citenamefont
  {Mohanty}, \citenamefont {Persson}, \citenamefont {Arvanitis}, \citenamefont
  {Temst},\ and\ \citenamefont {Haesendonck}}]{Mohanty2013}%
  \BibitemOpen
  \bibfield  {author} {\bibinfo {author} {\bibfnamefont {J}~\bibnamefont
  {Mohanty}}, \bibinfo {author} {\bibfnamefont {A}~\bibnamefont {Persson}},
  \bibinfo {author} {\bibfnamefont {D}~\bibnamefont {Arvanitis}}, \bibinfo
  {author} {\bibfnamefont {K}~\bibnamefont {Temst}}, \ and\ \bibinfo {author}
  {\bibfnamefont {C~Van}\ \bibnamefont {Haesendonck}},\ }\bibfield  {title}
  {\enquote {\bibinfo {title} {Direct observation of frozen moments in the
  {NiFe}/{FeMn} exchange bias system},}\ }\href {\doibase
  10.1088/1367-2630/15/3/033016} {\bibfield  {journal} {\bibinfo  {journal}
  {New Journal of Physics}\ }\textbf {\bibinfo {volume} {15}},\ \bibinfo
  {pages} {033016} (\bibinfo {year} {2013})}\BibitemShut {NoStop}%
\bibitem [{\citenamefont {Antel}\ \emph {et~al.}(1999)\citenamefont {Antel},
  \citenamefont {Perjeru},\ and\ \citenamefont {Harp}}]{Antel1999}%
  \BibitemOpen
  \bibfield  {author} {\bibinfo {author} {\bibfnamefont {W.~J.}\ \bibnamefont
  {Antel}}, \bibinfo {author} {\bibfnamefont {F.}~\bibnamefont {Perjeru}}, \
  and\ \bibinfo {author} {\bibfnamefont {G.~R.}\ \bibnamefont {Harp}},\
  }\bibfield  {title} {\enquote {\bibinfo {title} {Spin structure at the
  interface of exchange biased {FeMn/Co} bilayers},}\ }\href {\doibase
  10.1103/physrevlett.83.1439} {\bibfield  {journal} {\bibinfo  {journal}
  {Physical Review Letters}\ }\textbf {\bibinfo {volume} {83}},\ \bibinfo
  {pages} {1439--1442} (\bibinfo {year} {1999})}\BibitemShut {NoStop}%
\bibitem [{\citenamefont {O'Handley}(2000)}]{OHandley2000}%
  \BibitemOpen
  \bibfield  {author} {\bibinfo {author} {\bibfnamefont {R.~C.}\ \bibnamefont
  {O'Handley}},\ }\href@noop {} {\emph {\bibinfo {title} {Modern magnetic
  materials : principles and applications}}}\ (\bibinfo  {publisher}
  {Wiley-IEEE Press, New York},\ \bibinfo {year} {2000})\BibitemShut {NoStop}%
\bibitem [{\citenamefont {Nogu{\'{e}}s}\ \emph {et~al.}(2005)\citenamefont
  {Nogu{\'{e}}s}, \citenamefont {Sort}, \citenamefont {Langlais}, \citenamefont
  {Skumryev}, \citenamefont {Suri{\~{n}}ach}, \citenamefont {Mu{\~{n}}oz},\
  and\ \citenamefont {Bar{\'{o}}}}]{Nogues2005}%
  \BibitemOpen
  \bibfield  {author} {\bibinfo {author} {\bibfnamefont {J.}~\bibnamefont
  {Nogu{\'{e}}s}}, \bibinfo {author} {\bibfnamefont {J.}~\bibnamefont {Sort}},
  \bibinfo {author} {\bibfnamefont {V.}~\bibnamefont {Langlais}}, \bibinfo
  {author} {\bibfnamefont {V.}~\bibnamefont {Skumryev}}, \bibinfo {author}
  {\bibfnamefont {S.}~\bibnamefont {Suri{\~{n}}ach}}, \bibinfo {author}
  {\bibfnamefont {J.S.}\ \bibnamefont {Mu{\~{n}}oz}}, \ and\ \bibinfo {author}
  {\bibfnamefont {M.D.}\ \bibnamefont {Bar{\'{o}}}},\ }\bibfield  {title}
  {\enquote {\bibinfo {title} {Exchange bias in nanostructures},}\ }\href
  {\doibase 10.1016/j.physrep.2005.08.004} {\bibfield  {journal} {\bibinfo
  {journal} {Physics Reports}\ }\textbf {\bibinfo {volume} {422}},\ \bibinfo
  {pages} {65--117} (\bibinfo {year} {2005})}\BibitemShut {NoStop}%
\bibitem [{\citenamefont {Leighton}\ \emph {et~al.}(2002)\citenamefont
  {Leighton}, \citenamefont {Fitzsimmons}, \citenamefont {Hoffmann},
  \citenamefont {Dura}, \citenamefont {Majkrzak}, \citenamefont {Lund},\ and\
  \citenamefont {Schuller}}]{Leighton2002}%
  \BibitemOpen
  \bibfield  {author} {\bibinfo {author} {\bibfnamefont {C.}~\bibnamefont
  {Leighton}}, \bibinfo {author} {\bibfnamefont {M.~R.}\ \bibnamefont
  {Fitzsimmons}}, \bibinfo {author} {\bibfnamefont {A.}~\bibnamefont
  {Hoffmann}}, \bibinfo {author} {\bibfnamefont {J.}~\bibnamefont {Dura}},
  \bibinfo {author} {\bibfnamefont {C.~F.}\ \bibnamefont {Majkrzak}}, \bibinfo
  {author} {\bibfnamefont {M.~S.}\ \bibnamefont {Lund}}, \ and\ \bibinfo
  {author} {\bibfnamefont {Ivan~K.}\ \bibnamefont {Schuller}},\ }\bibfield
  {title} {\enquote {\bibinfo {title} {Thickness-dependent coercive mechanisms
  in exchange-biased bilayers},}\ }\href {\doibase 10.1103/physrevb.65.064403}
  {\bibfield  {journal} {\bibinfo  {journal} {Physical Review B}\ }\textbf
  {\bibinfo {volume} {65}},\ \bibinfo {pages} {644031--644037} (\bibinfo {year}
  {2002})}\BibitemShut {NoStop}%
\bibitem [{\citenamefont {Merodio}\ \emph {et~al.}(2014)\citenamefont
  {Merodio}, \citenamefont {Ghosh}, \citenamefont {Lemonias}, \citenamefont
  {Gautier}, \citenamefont {Ebels}, \citenamefont {Chshiev}, \citenamefont
  {B{\'{e}}a}, \citenamefont {Baltz},\ and\ \citenamefont
  {Bailey}}]{Merodio2014}%
  \BibitemOpen
  \bibfield  {author} {\bibinfo {author} {\bibfnamefont {P.}~\bibnamefont
  {Merodio}}, \bibinfo {author} {\bibfnamefont {A.}~\bibnamefont {Ghosh}},
  \bibinfo {author} {\bibfnamefont {C.}~\bibnamefont {Lemonias}}, \bibinfo
  {author} {\bibfnamefont {E.}~\bibnamefont {Gautier}}, \bibinfo {author}
  {\bibfnamefont {U.}~\bibnamefont {Ebels}}, \bibinfo {author} {\bibfnamefont
  {M.}~\bibnamefont {Chshiev}}, \bibinfo {author} {\bibfnamefont
  {H.}~\bibnamefont {B{\'{e}}a}}, \bibinfo {author} {\bibfnamefont
  {V.}~\bibnamefont {Baltz}}, \ and\ \bibinfo {author} {\bibfnamefont {W.~E.}\
  \bibnamefont {Bailey}},\ }\bibfield  {title} {\enquote {\bibinfo {title}
  {Penetration depth and absorption mechanisms of spin currents in
  {Ir$_{20}$Mn$_{80}$} and {Fe$_{50}$Mn$_{50}$} polycrystalline films by
  ferromagnetic resonance and spin pumping},}\ }\href {\doibase
  10.1063/1.4862971} {\bibfield  {journal} {\bibinfo  {journal} {Applied
  Physics Letters}\ }\textbf {\bibinfo {volume} {104}},\ \bibinfo {pages}
  {032406} (\bibinfo {year} {2014})}\BibitemShut {NoStop}%
\bibitem [{\citenamefont {Hernando}\ \emph {et~al.}(1995)\citenamefont
  {Hernando}, \citenamefont {Navarro},\ and\ \citenamefont
  {Gorr{\'{\i}}a}}]{Hernando1995}%
  \BibitemOpen
  \bibfield  {author} {\bibinfo {author} {\bibfnamefont {A.}~\bibnamefont
  {Hernando}}, \bibinfo {author} {\bibfnamefont {I.}~\bibnamefont {Navarro}}, \
  and\ \bibinfo {author} {\bibfnamefont {P.}~\bibnamefont {Gorr{\'{\i}}a}},\
  }\bibfield  {title} {\enquote {\bibinfo {title} {Iron exchange-field
  penetration into the amorphous interphase of nanocrystalline materials},}\
  }\href {\doibase 10.1103/physrevb.51.3281} {\bibfield  {journal} {\bibinfo
  {journal} {Physical Review B}\ }\textbf {\bibinfo {volume} {51}},\ \bibinfo
  {pages} {3281--3284} (\bibinfo {year} {1995})}\BibitemShut {NoStop}%
\bibitem [{\citenamefont {Navarro}\ \emph {et~al.}(1996)\citenamefont
  {Navarro}, \citenamefont {Ortu{\~{n}}o},\ and\ \citenamefont
  {Hernando}}]{Navarro1996}%
  \BibitemOpen
  \bibfield  {author} {\bibinfo {author} {\bibfnamefont {I.}~\bibnamefont
  {Navarro}}, \bibinfo {author} {\bibfnamefont {M.}~\bibnamefont
  {Ortu{\~{n}}o}}, \ and\ \bibinfo {author} {\bibfnamefont {A.}~\bibnamefont
  {Hernando}},\ }\bibfield  {title} {\enquote {\bibinfo {title} {Ferromagnetic
  interactions in nanostructured systems with two different curie
  temperatures},}\ }\href {\doibase 10.1103/physrevb.53.11656} {\bibfield
  {journal} {\bibinfo  {journal} {Physical Review B}\ }\textbf {\bibinfo
  {volume} {53}},\ \bibinfo {pages} {11656--11660} (\bibinfo {year}
  {1996})}\BibitemShut {NoStop}%
\bibitem [{\citenamefont {Lenz}\ \emph {et~al.}(2007)\citenamefont {Lenz},
  \citenamefont {Zander},\ and\ \citenamefont {Kuch}}]{Lenz2007}%
  \BibitemOpen
  \bibfield  {author} {\bibinfo {author} {\bibfnamefont {K.}~\bibnamefont
  {Lenz}}, \bibinfo {author} {\bibfnamefont {S.}~\bibnamefont {Zander}}, \ and\
  \bibinfo {author} {\bibfnamefont {W.}~\bibnamefont {Kuch}},\ }\bibfield
  {title} {\enquote {\bibinfo {title} {Magnetic proximity effects in
  antiferromagnet/ferromagnet bilayers: The impact on the {N{\'{e}}el}
  temperature},}\ }\href {\doibase 10.1103/physrevlett.98.237201} {\bibfield
  {journal} {\bibinfo  {journal} {Physical Review Letters}\ }\textbf {\bibinfo
  {volume} {98}},\ \bibinfo {pages} {237201} (\bibinfo {year}
  {2007})}\BibitemShut {NoStop}%
\bibitem [{\citenamefont {Saglam}\ \emph {et~al.}(2016)\citenamefont {Saglam},
  \citenamefont {Zhang}, \citenamefont {Jungfleisch}, \citenamefont {Sklenar},
  \citenamefont {Pearson}, \citenamefont {Ketterson},\ and\ \citenamefont
  {Hoffmann}}]{Saglam2016}%
  \BibitemOpen
  \bibfield  {author} {\bibinfo {author} {\bibfnamefont {H.}~\bibnamefont
  {Saglam}}, \bibinfo {author} {\bibfnamefont {W.}~\bibnamefont {Zhang}},
  \bibinfo {author} {\bibfnamefont {M.~B.}\ \bibnamefont {Jungfleisch}},
  \bibinfo {author} {\bibfnamefont {J.}~\bibnamefont {Sklenar}}, \bibinfo
  {author} {\bibfnamefont {J.~E.}\ \bibnamefont {Pearson}}, \bibinfo {author}
  {\bibfnamefont {J.~B.}\ \bibnamefont {Ketterson}}, \ and\ \bibinfo {author}
  {\bibfnamefont {A.}~\bibnamefont {Hoffmann}},\ }\bibfield  {title} {\enquote
  {\bibinfo {title} {Spin transport through the metallic antiferromagnet
  {FeMn}},}\ }\href {\doibase 10.1103/physrevb.94.140412} {\bibfield  {journal}
  {\bibinfo  {journal} {Physical Review B}\ }\textbf {\bibinfo {volume} {94}},\
  \bibinfo {pages} {140412} (\bibinfo {year} {2016})}\BibitemShut {NoStop}%
\bibitem [{\citenamefont {Ekholm}\ and\ \citenamefont
  {Abrikosov}(2011)}]{Ekholm2011}%
  \BibitemOpen
  \bibfield  {author} {\bibinfo {author} {\bibfnamefont {M.}~\bibnamefont
  {Ekholm}}\ and\ \bibinfo {author} {\bibfnamefont {I.~A.}\ \bibnamefont
  {Abrikosov}},\ }\bibfield  {title} {\enquote {\bibinfo {title} {Structural
  and magnetic ground-state properties of $\gamma$-{FeMn} alloys from ab initio
  calculations},}\ }\href {\doibase 10.1103/physrevb.84.104423} {\bibfield
  {journal} {\bibinfo  {journal} {Physical Review B}\ }\textbf {\bibinfo
  {volume} {84}},\ \bibinfo {pages} {104423} (\bibinfo {year}
  {2011})}\BibitemShut {NoStop}%
\bibitem [{\citenamefont {Polishchuk}\ \emph
  {et~al.}(2021{\natexlab{a}})\citenamefont {Polishchuk}, \citenamefont
  {Tykhonenko-Polishchuk}, \citenamefont {Lytvynenko}, \citenamefont {Rostas},
  \citenamefont {Gomonay},\ and\ \citenamefont {Korenivski}}]{Polishchuk2021}%
  \BibitemOpen
  \bibfield  {author} {\bibinfo {author} {\bibfnamefont
  {D.{\hspace{0.167em}}M.}\ \bibnamefont {Polishchuk}}, \bibinfo {author}
  {\bibfnamefont {Yu.{\hspace{0.167em}}O.}\ \bibnamefont
  {Tykhonenko-Polishchuk}}, \bibinfo {author} {\bibfnamefont
  {Ya.{\hspace{0.167em}}M.}\ \bibnamefont {Lytvynenko}}, \bibinfo {author}
  {\bibfnamefont {A.{\hspace{0.167em}}M.}\ \bibnamefont {Rostas}}, \bibinfo
  {author} {\bibfnamefont {O.{\hspace{0.167em}}V.}\ \bibnamefont {Gomonay}}, \
  and\ \bibinfo {author} {\bibfnamefont {V.}~\bibnamefont {Korenivski}},\
  }\bibfield  {title} {\enquote {\bibinfo {title} {Thermal gating of magnon
  exchange in magnetic multilayers with antiferromagnetic spacers},}\ }\href
  {\doibase 10.1103/physrevlett.126.227203} {\bibfield  {journal} {\bibinfo
  {journal} {Physical Review Letters}\ }\textbf {\bibinfo {volume} {126}},\
  \bibinfo {pages} {227203} (\bibinfo {year} {2021}{\natexlab{a}})}\BibitemShut
  {NoStop}%
\bibitem [{\citenamefont {Nederpel}\ and\ \citenamefont
  {Martens}(1985)}]{Nederpel1985}%
  \BibitemOpen
  \bibfield  {author} {\bibinfo {author} {\bibfnamefont {P.~Q.~J.}\
  \bibnamefont {Nederpel}}\ and\ \bibinfo {author} {\bibfnamefont {J.~W.~D.}\
  \bibnamefont {Martens}},\ }\bibfield  {title} {\enquote {\bibinfo {title}
  {Magneto‐optical ellipsometer},}\ }\href {\doibase 10.1063/1.1138206}
  {\bibfield  {journal} {\bibinfo  {journal} {Review of Scientific
  Instruments}\ }\textbf {\bibinfo {volume} {56}},\ \bibinfo {pages} {687--690}
  (\bibinfo {year} {1985})},\ \Eprint
  {http://arxiv.org/abs/https://doi.org/10.1063/1.1138206}
  {https://doi.org/10.1063/1.1138206} \BibitemShut {NoStop}%
\bibitem [{\citenamefont {Polisetty}\ \emph {et~al.}(2008)\citenamefont
  {Polisetty}, \citenamefont {Scheffler}, \citenamefont {Sahoo}, \citenamefont
  {Wang}, \citenamefont {Mukherjee}, \citenamefont {He},\ and\ \citenamefont
  {Binek}}]{Polisetty2008}%
  \BibitemOpen
  \bibfield  {author} {\bibinfo {author} {\bibfnamefont {S.}~\bibnamefont
  {Polisetty}}, \bibinfo {author} {\bibfnamefont {J.}~\bibnamefont
  {Scheffler}}, \bibinfo {author} {\bibfnamefont {S.}~\bibnamefont {Sahoo}},
  \bibinfo {author} {\bibfnamefont {Yi}~\bibnamefont {Wang}}, \bibinfo {author}
  {\bibfnamefont {T.}~\bibnamefont {Mukherjee}}, \bibinfo {author}
  {\bibfnamefont {Xi}~\bibnamefont {He}}, \ and\ \bibinfo {author}
  {\bibfnamefont {Ch.}\ \bibnamefont {Binek}},\ }\bibfield  {title} {\enquote
  {\bibinfo {title} {Optimization of magneto-optical kerr setup: Analyzing
  experimental assemblies using jones matrix formalism},}\ }\href {\doibase
  10.1063/1.2932445} {\bibfield  {journal} {\bibinfo  {journal} {Review of
  Scientific Instruments}\ }\textbf {\bibinfo {volume} {79}},\ \bibinfo {pages}
  {055107} (\bibinfo {year} {2008})},\ \Eprint
  {http://arxiv.org/abs/https://doi.org/10.1063/1.2932445}
  {https://doi.org/10.1063/1.2932445} \BibitemShut {NoStop}%
\bibitem [{\citenamefont {Kravets}\ \emph {et~al.}(2014)\citenamefont
  {Kravets}, \citenamefont {Dzhezherya}, \citenamefont {Tovstolytkin},
  \citenamefont {Kozak}, \citenamefont {Gryshchuk}, \citenamefont {Savina},
  \citenamefont {Pashchenko}, \citenamefont {Gnatchenko}, \citenamefont
  {Koop},\ and\ \citenamefont {Korenivski}}]{Kravets_2014}%
  \BibitemOpen
  \bibfield  {author} {\bibinfo {author} {\bibfnamefont {A.~F.}\ \bibnamefont
  {Kravets}}, \bibinfo {author} {\bibfnamefont {Yu.~I.}\ \bibnamefont
  {Dzhezherya}}, \bibinfo {author} {\bibfnamefont {A.~I.}\ \bibnamefont
  {Tovstolytkin}}, \bibinfo {author} {\bibfnamefont {I.~M.}\ \bibnamefont
  {Kozak}}, \bibinfo {author} {\bibfnamefont {A.}~\bibnamefont {Gryshchuk}},
  \bibinfo {author} {\bibfnamefont {Yu.~O.}\ \bibnamefont {Savina}}, \bibinfo
  {author} {\bibfnamefont {V.~A.}\ \bibnamefont {Pashchenko}}, \bibinfo
  {author} {\bibfnamefont {S.~L.}\ \bibnamefont {Gnatchenko}}, \bibinfo
  {author} {\bibfnamefont {B.}~\bibnamefont {Koop}}, \ and\ \bibinfo {author}
  {\bibfnamefont {V.}~\bibnamefont {Korenivski}},\ }\bibfield  {title}
  {\enquote {\bibinfo {title} {Synthetic ferrimagnets with thermomagnetic
  switching},}\ }\href {\doibase https://doi.org/10.1103/PhysRevB.90.104427}
  {\bibfield  {journal} {\bibinfo  {journal} {Physical Review B}\ }\textbf
  {\bibinfo {volume} {90}},\ \bibinfo {pages} {104427} (\bibinfo {year}
  {2014})}\BibitemShut {NoStop}%
\bibitem [{\citenamefont {Kravets}\ \emph {et~al.}(2015)\citenamefont
  {Kravets}, \citenamefont {Tovstolytkin}, \citenamefont {Dzhezherya},
  \citenamefont {Polishchuk}, \citenamefont {Kozak},\ and\ \citenamefont
  {Korenivski}}]{Kravets_2015}%
  \BibitemOpen
  \bibfield  {author} {\bibinfo {author} {\bibfnamefont {A~F}\ \bibnamefont
  {Kravets}}, \bibinfo {author} {\bibfnamefont {A~I}\ \bibnamefont
  {Tovstolytkin}}, \bibinfo {author} {\bibfnamefont {Yu~I}\ \bibnamefont
  {Dzhezherya}}, \bibinfo {author} {\bibfnamefont {D~M}\ \bibnamefont
  {Polishchuk}}, \bibinfo {author} {\bibfnamefont {I~M}\ \bibnamefont {Kozak}},
  \ and\ \bibinfo {author} {\bibfnamefont {V}~\bibnamefont {Korenivski}},\
  }\bibfield  {title} {\enquote {\bibinfo {title} {Spin dynamics in a
  {Curie}-switch},}\ }\href {\doibase
  https://doi.org/10.1088/0953-8984/27/44/446003} {\bibfield  {journal}
  {\bibinfo  {journal} {Journal of Physics: Condensed Matter}\ }\textbf
  {\bibinfo {volume} {27}},\ \bibinfo {pages} {446003} (\bibinfo {year}
  {2015})}\BibitemShut {NoStop}%
\bibitem [{\citenamefont {Kravets}\ \emph {et~al.}(2016)\citenamefont
  {Kravets}, \citenamefont {Polishchuk}, \citenamefont {Dzhezherya},
  \citenamefont {Tovstolytkin}, \citenamefont {Golub},\ and\ \citenamefont
  {Korenivski}}]{Kravets_2016}%
  \BibitemOpen
  \bibfield  {author} {\bibinfo {author} {\bibfnamefont {A.~F.}\ \bibnamefont
  {Kravets}}, \bibinfo {author} {\bibfnamefont {D.~M.}\ \bibnamefont
  {Polishchuk}}, \bibinfo {author} {\bibfnamefont {Yu.~I.}\ \bibnamefont
  {Dzhezherya}}, \bibinfo {author} {\bibfnamefont {A.~I.}\ \bibnamefont
  {Tovstolytkin}}, \bibinfo {author} {\bibfnamefont {V.~O.}\ \bibnamefont
  {Golub}}, \ and\ \bibinfo {author} {\bibfnamefont {V.}~\bibnamefont
  {Korenivski}},\ }\bibfield  {title} {\enquote {\bibinfo {title} {Anisotropic
  magnetization relaxation in ferromagnetic multilayers with variable
  interlayer exchange coupling},}\ }\href {\doibase
  https://doi.org/10.1103/PhysRevB.94.064429} {\bibfield  {journal} {\bibinfo
  {journal} {Physical Review B}\ }\textbf {\bibinfo {volume} {94}},\ \bibinfo
  {pages} {064429} (\bibinfo {year} {2016})}\BibitemShut {NoStop}%
\bibitem [{\citenamefont {Polishchuk}\ \emph {et~al.}(2018)\citenamefont
  {Polishchuk}, \citenamefont {Tykhonenko-Polishchuk}, \citenamefont
  {Holmgren}, \citenamefont {Kravets}, \citenamefont {Tovstolytkin},\ and\
  \citenamefont {Korenivski}}]{Polishchuk2018}%
  \BibitemOpen
  \bibfield  {author} {\bibinfo {author} {\bibfnamefont {D.~M.}\ \bibnamefont
  {Polishchuk}}, \bibinfo {author} {\bibfnamefont {Yu.~O.}\ \bibnamefont
  {Tykhonenko-Polishchuk}}, \bibinfo {author} {\bibfnamefont {E.}~\bibnamefont
  {Holmgren}}, \bibinfo {author} {\bibfnamefont {A.~F.}\ \bibnamefont
  {Kravets}}, \bibinfo {author} {\bibfnamefont {A.~I.}\ \bibnamefont
  {Tovstolytkin}}, \ and\ \bibinfo {author} {\bibfnamefont {V.}~\bibnamefont
  {Korenivski}},\ }\bibfield  {title} {\enquote {\bibinfo {title} {Giant
  magnetocaloric effect driven by indirect exchange in magnetic multilayers},}\
  }\href {\doibase 10.1103/physrevmaterials.2.114402} {\bibfield  {journal}
  {\bibinfo  {journal} {Physical Review Materials}\ }\textbf {\bibinfo {volume}
  {2}},\ \bibinfo {pages} {114402} (\bibinfo {year} {2018})}\BibitemShut
  {NoStop}%
\bibitem [{\citenamefont {Kaya}\ \emph {et~al.}(2013)\citenamefont {Kaya},
  \citenamefont {Lapa}, \citenamefont {Jayathilaka}, \citenamefont {Kirby},
  \citenamefont {Miller},\ and\ \citenamefont {Roshchin}}]{Kaya2013}%
  \BibitemOpen
  \bibfield  {author} {\bibinfo {author} {\bibfnamefont {Dogan}\ \bibnamefont
  {Kaya}}, \bibinfo {author} {\bibfnamefont {Pavel~N.}\ \bibnamefont {Lapa}},
  \bibinfo {author} {\bibfnamefont {Priyanga}\ \bibnamefont {Jayathilaka}},
  \bibinfo {author} {\bibfnamefont {Hillary}\ \bibnamefont {Kirby}}, \bibinfo
  {author} {\bibfnamefont {Casey~W.}\ \bibnamefont {Miller}}, \ and\ \bibinfo
  {author} {\bibfnamefont {Igor~V.}\ \bibnamefont {Roshchin}},\ }\bibfield
  {title} {\enquote {\bibinfo {title} {Controlling exchange bias in {FeMn} with
  {Cu}},}\ }\href {\doibase 10.1063/1.4798310} {\bibfield  {journal} {\bibinfo
  {journal} {Journal of Applied Physics}\ }\textbf {\bibinfo {volume} {113}},\
  \bibinfo {pages} {17D717} (\bibinfo {year} {2013})}\BibitemShut {NoStop}%
\bibitem [{\citenamefont {Schmitz}\ \emph {et~al.}(2010)\citenamefont
  {Schmitz}, \citenamefont {Schierle}, \citenamefont {Darowski}, \citenamefont
  {Maletta}, \citenamefont {Weschke},\ and\ \citenamefont
  {Gruyters}}]{Schmitz2010}%
  \BibitemOpen
  \bibfield  {author} {\bibinfo {author} {\bibfnamefont {D.}~\bibnamefont
  {Schmitz}}, \bibinfo {author} {\bibfnamefont {E.}~\bibnamefont {Schierle}},
  \bibinfo {author} {\bibfnamefont {N.}~\bibnamefont {Darowski}}, \bibinfo
  {author} {\bibfnamefont {H.}~\bibnamefont {Maletta}}, \bibinfo {author}
  {\bibfnamefont {E.}~\bibnamefont {Weschke}}, \ and\ \bibinfo {author}
  {\bibfnamefont {M.}~\bibnamefont {Gruyters}},\ }\bibfield  {title} {\enquote
  {\bibinfo {title} {Unidirectional behavior of uncompensated {Fe} orbital
  moments in exchange-biased {Co/FeMn/Cu(001)}},}\ }\href {\doibase
  10.1103/physrevb.81.224422} {\bibfield  {journal} {\bibinfo  {journal}
  {Physical Review B}\ }\textbf {\bibinfo {volume} {81}},\ \bibinfo {pages}
  {224422} (\bibinfo {year} {2010})}\BibitemShut {NoStop}%
\bibitem [{\citenamefont {Polishchuk}\ \emph
  {et~al.}(2021{\natexlab{b}})\citenamefont {Polishchuk}, \citenamefont
  {Polek}, \citenamefont {Borynskyi}, \citenamefont {Kravets}, \citenamefont
  {Tovstolytkin},\ and\ \citenamefont {Korenivski}}]{Polishchuk2021a}%
  \BibitemOpen
  \bibfield  {author} {\bibinfo {author} {\bibfnamefont {D~M}\ \bibnamefont
  {Polishchuk}}, \bibinfo {author} {\bibfnamefont {T~I}\ \bibnamefont {Polek}},
  \bibinfo {author} {\bibfnamefont {V~Yu}\ \bibnamefont {Borynskyi}}, \bibinfo
  {author} {\bibfnamefont {A~F}\ \bibnamefont {Kravets}}, \bibinfo {author}
  {\bibfnamefont {A~I}\ \bibnamefont {Tovstolytkin}}, \ and\ \bibinfo {author}
  {\bibfnamefont {V}~\bibnamefont {Korenivski}},\ }\bibfield  {title} {\enquote
  {\bibinfo {title} {Isotropic {FMR} frequency enhancement in thin {Py/FeMn}
  bilayers under strong magnetic proximity effect},}\ }\href {\doibase
  10.1088/1361-6463/abfe39} {\bibfield  {journal} {\bibinfo  {journal} {Journal
  of Physics D: Applied Physics}\ }\textbf {\bibinfo {volume} {54}},\ \bibinfo
  {pages} {305003} (\bibinfo {year} {2021}{\natexlab{b}})}\BibitemShut
  {NoStop}%
\bibitem [{\citenamefont {Kittel}(1948)}]{Kittel1948}%
  \BibitemOpen
  \bibfield  {author} {\bibinfo {author} {\bibfnamefont {Charles}\ \bibnamefont
  {Kittel}},\ }\bibfield  {title} {\enquote {\bibinfo {title} {On the theory of
  ferromagnetic resonance absorption},}\ }\href {\doibase
  10.1103/physrev.73.155} {\bibfield  {journal} {\bibinfo  {journal} {Physical
  Review}\ }\textbf {\bibinfo {volume} {73}},\ \bibinfo {pages} {155--161}
  (\bibinfo {year} {1948})}\BibitemShut {NoStop}%
\bibitem [{\citenamefont {Kuch}\ \emph {et~al.}(2004)\citenamefont {Kuch},
  \citenamefont {Chelaru}, \citenamefont {Offi}, \citenamefont {Wang},
  \citenamefont {Kotsugi},\ and\ \citenamefont
  {Kirschner}}]{Kuch:PhysRevLett.92.017201}%
  \BibitemOpen
  \bibfield  {author} {\bibinfo {author} {\bibfnamefont {Wolfgang}\
  \bibnamefont {Kuch}}, \bibinfo {author} {\bibfnamefont {Liviu~I.}\
  \bibnamefont {Chelaru}}, \bibinfo {author} {\bibfnamefont {Francesco}\
  \bibnamefont {Offi}}, \bibinfo {author} {\bibfnamefont {Jing}\ \bibnamefont
  {Wang}}, \bibinfo {author} {\bibfnamefont {Masato}\ \bibnamefont {Kotsugi}},
  \ and\ \bibinfo {author} {\bibfnamefont {J\"urgen}\ \bibnamefont
  {Kirschner}},\ }\bibfield  {title} {\enquote {\bibinfo {title}
  {Three-dimensional noncollinear antiferromagnetic order in single-crystalline
  {FeMn} ultrathin films},}\ }\href {\doibase 10.1103/PhysRevLett.92.017201}
  {\bibfield  {journal} {\bibinfo  {journal} {Phys. Rev. Lett.}\ }\textbf
  {\bibinfo {volume} {92}},\ \bibinfo {pages} {017201} (\bibinfo {year}
  {2004})}\BibitemShut {NoStop}%
\bibitem [{Note1()}]{Note1}%
  \BibitemOpen
  \bibinfo {note} {In an ultrathin AF spacer, the magnetic sublattices are no
  longer equivalent due to the presence of the ferromagnetic exchange at the
  interfaces, so Eq.~\protect \textup {\hbox {\mathsurround \z@ \protect
  \normalfont (\ignorespaces \ref {eq_interaction_dimension}\unskip
  \@@italiccorr )}} should be modified to include the coupling with both the
  N\'eel vector and the proximity-induced AF magnetization $\protect \mathbf
  {M}_\protect \mathrm {AF}$.}\BibitemShut {Stop}%
\bibitem [{\citenamefont {Polishchuk}\ \emph {et~al.}(2023)\citenamefont
  {Polishchuk}, \citenamefont {Persson}, \citenamefont {Kulyk}, \citenamefont
  {Baglioni}, \citenamefont {Ivanov},\ and\ \citenamefont
  {Korenivski}}]{Polishchuk:10.1063_5.0133125}%
  \BibitemOpen
  \bibfield  {author} {\bibinfo {author} {\bibfnamefont {D.~M.}\ \bibnamefont
  {Polishchuk}}, \bibinfo {author} {\bibfnamefont {M.}~\bibnamefont {Persson}},
  \bibinfo {author} {\bibfnamefont {M.~M.}\ \bibnamefont {Kulyk}}, \bibinfo
  {author} {\bibfnamefont {G.}~\bibnamefont {Baglioni}}, \bibinfo {author}
  {\bibfnamefont {B.~A.}\ \bibnamefont {Ivanov}}, \ and\ \bibinfo {author}
  {\bibfnamefont {V.}~\bibnamefont {Korenivski}},\ }\bibfield  {title}
  {\enquote {\bibinfo {title} {Oscillatory exchange bias controlled by {RKKY}
  in magnetic multilayers},}\ }\href {\doibase 10.1063/5.0133125} {\bibfield
  {journal} {\bibinfo  {journal} {Applied Physics Letters}\ }\textbf {\bibinfo
  {volume} {122}},\ \bibinfo {pages} {062405} (\bibinfo {year}
  {2023})}\BibitemShut {NoStop}%
\bibitem [{\citenamefont {Brown}(1963)}]{Brown1963}%
  \BibitemOpen
  \bibfield  {author} {\bibinfo {author} {\bibfnamefont {William~Fuller}\
  \bibnamefont {Brown}},\ }\bibfield  {title} {\enquote {\bibinfo {title}
  {{Thermal Fluctuations of a Single-Domain Particle}},}\ }\href {\doibase
  10.1103/PhysRev.130.1677} {\bibfield  {journal} {\bibinfo  {journal}
  {Physical Review}\ }\textbf {\bibinfo {volume} {130}},\ \bibinfo {pages}
  {1677--1686} (\bibinfo {year} {1963})},\ \Eprint
  {http://arxiv.org/abs/arXiv:1011.1669v3} {arXiv:arXiv:1011.1669v3}
  \BibitemShut {NoStop}%
\bibitem [{\citenamefont {Peng}\ and\ \citenamefont
  {Richter}(2004)}]{Peng2004}%
  \BibitemOpen
  \bibfield  {author} {\bibinfo {author} {\bibfnamefont {Qingzhi}\ \bibnamefont
  {Peng}}\ and\ \bibinfo {author} {\bibfnamefont {Hans~J.}\ \bibnamefont
  {Richter}},\ }\bibfield  {title} {\enquote {\bibinfo {title} {{Field sweep
  rate dependence of media dynamic coercivity}},}\ }\href {\doibase
  10.1109/TMAG.2004.829021} {\bibfield  {journal} {\bibinfo  {journal} {IEEE
  Transactions on Magnetics}\ }\textbf {\bibinfo {volume} {40}},\ \bibinfo
  {pages} {2446--2448} (\bibinfo {year} {2004})}\BibitemShut {NoStop}%
\bibitem [{\citenamefont {Gomonay}\ and\ \citenamefont
  {Loktev}(2014)}]{Gomonay2014b}%
  \BibitemOpen
  \bibfield  {author} {\bibinfo {author} {\bibfnamefont {E.~V.}\ \bibnamefont
  {Gomonay}}\ and\ \bibinfo {author} {\bibfnamefont {V.~M.}\ \bibnamefont
  {Loktev}},\ }\bibfield  {title} {\enquote {\bibinfo {title} {{Spintronics of
  antiferromagnetic systems (Review Article)}},}\ }\href {\doibase
  10.1063/1.4862467} {\bibfield  {journal} {\bibinfo  {journal} {Low Temp.
  Phys.}\ }\textbf {\bibinfo {volume} {40}},\ \bibinfo {pages} {17--35}
  (\bibinfo {year} {2014})}\BibitemShut {NoStop}%
\bibitem [{\citenamefont {Lifshitz}(1955)}]{Lifshitz1955}%
  \BibitemOpen
  \bibfield  {author} {\bibinfo {author} {\bibfnamefont {E.M.}\ \bibnamefont
  {Lifshitz}},\ }\bibfield  {title} {\enquote {\bibinfo {title} {{The theory of
  molecular attractive forces between solids}},}\ }\href@noop {} {\bibfield
  {journal} {\bibinfo  {journal} {J. Exper. Theoret. Phys.}\ }\textbf {\bibinfo
  {volume} {29}},\ \bibinfo {pages} {94--110} (\bibinfo {year}
  {1955})}\BibitemShut {NoStop}%
\bibitem [{Note2()}]{Note2}%
  \BibitemOpen
  \bibinfo {note} {~The magnon spectra of an antiferromagnet consist of two
  branches with different polarization. However, in the present model we
  consider only one branch, which hybridises with the ferromagnetic magnons at
  the interfaces.}\BibitemShut {Stop}%
\end{thebibliography}%

\end{document}